\documentclass[%
 amsmath,amssymb,
 aps,
 prfluids,
floatfix,
]{revtex4-2}

\usepackage{natbib}

\usepackage{graphicx,color}
\usepackage[normalem]{ulem}
\usepackage{amsmath}
\usepackage{tabularx}
\usepackage{graphicx} 
\usepackage{appendix}
\usepackage{bm}
	
\usepackage{pdfpages}

\makeatletter
\AtBeginDocument{\let\LS@rot\@undefined}
\makeatother

\definecolor{colortodo}{RGB}{0,0,0}

\definecolor{colortodo2}{RGB}{0,0,255}
\definecolor{colortodo2}{RGB}{0,0,0}

\newcommand{\blue}[1]{{\color{colortodo2}#1}}

\definecolor{colortodo3}{RGB}{255,255,255}

\begin{document}

\title{Self-propulsion of floating ice blocks caused by melting in water}

\author{Michael Berhanu$^\star$, Amit Dawadi$^\dagger$, Martin Chaigne$^\star$, J\'er\^ome Jovet$^\S$ and Arshad Kudrolli$^\dagger$}
\affiliation{$^\star$ MSC, Université Paris Cité, CNRS (UMR 7057), F-75013 Paris, France}
\affiliation{$^\S$ Physics Department, Université Paris Cité, F-75013 Paris, France}
\affiliation{$^\dagger$ Department of Physics, Clark University, Worcester, Massachusetts}

\date{\today}

\begin{abstract}
We show that floating ice blocks with asymmetric shapes can self-propel with significant speeds due to buoyancy driven currents caused by melting. \blue{In water baths with temperatures between $10\,^\circ$C and $30\,^\circ$C, model right-angle ice wedges are found to move in the direction opposite to the gravity current which descends along the longest inclined side}. We describe the measured speed as a function of the length and angle of the inclined side, and the temperature of the bath in terms of a propulsion model which incorporates the cooling of the surrounding fluid by the melting of ice.  \blue{The heat pulled from the surrounding liquid by the melting ice block generates a thermal convection flow, leading to momentum exchange and 
a net propulsion force. The translation velocity is explained by balancing the propulsion force by drag.} We further show that the ice block moves robustly in a saltwater bath with ocean-like salinity and maintains the same direction of motion as in freshwater. A simplified model is further developed to describe the propulsion of asymmetric ice blocks in saltwater, incorporating the effects of rising meltwater and the sinking of the surrounding bath water due to cooling.  For sufficiently large temperature, we find that the cooling-induced sinking flow generates a stronger force than the upward flow from the meltwater. Consequently, the net propulsion force is in the same direction and nearly the same magnitude as that observed in freshwater. These findings suggest that melting-driven propulsion may be relevant to the motion of icebergs in sufficiently warm oceanic environments.
\end{abstract}

\maketitle

\section{Introduction}
The melting of icebergs floating in the ocean is often accompanied by buoyant convection flows~\cite{cenedese2023}, as local temperature and salinity variations modify the water density. Consequently, significant gravity driven currents occur below the water surface in the vicinity of icebergs. These currents carry momentum and can contribute to iceberg motion, in addition to contributions due to oceanic currents, wind, surface waves, and Coriolis force~\cite{Mountain1980,Bigg1997,Wagner2017,Marchenko2019}. {The idea that ice melting can lead to a propulsion effect relevant for iceberg was raised by Mercier, {\it et al.}~\cite{Mercier2014}} as a perspective to their work in which they demonstrated {slow directed motion} of a floating asymmetric solid with an embedded local heat source that generated thermal convection. {The same group had reported the self-propulsion of an asymmetric object in a density-stratified flow driven by diffusion~\cite{Allshouse2010}, but the observed velocities were several orders of magnitude smaller than those with heated blocks.} 

An additional source of energy is not required to create a heat flux and convection current for a melting block floating in a bath at a temperature different from the melting temperature. Previously, Dorbolo, {\it et al.}~\cite{Dorbolo2016} related the spinning of floating ice disks to the convection flow driven by  melting. No translation was reported because the disks were symmetric and were constrained to rotate by fixing the center position using magnets. Recently, it was demonstrated that a boat incorporating an inclined solute material like salt or sugar can propel rapidly due to the solutal convection flow driven by the dissolution~\cite{Chaigne2023}. We build on that study by investigating the case of asymmetric ice blocks as they melt in warm water. 
As in the case of dissolution, the convection flow generated by density variations due to a phase change from solid to liquid leads to self-propulsion, if the convection flow displays a forward-backward asymmetry. The flow then has a non-zero horizontal momentum component, which results in an oppositely directed reaction propulsion force on the body, that can be estimated using a momentum balance. 
\blue{However, melting-driven propulsion is not equivalent to dissolution-driven propulsion. Melting and dissolution are 
fundamentally different even though both are phase changes from solid to liquid~\cite{carpy2024fingerprints}.} 

Melting is driven by temperature transport and the Stefan condition at the solid-liquid interface results from an energy balance involving latent heat. By contrast, dissolution of salt or sugar in water is driven by solute transport and the  boundary condition corresponds to the solute mass conservation balance. Moreover, in the case of ice melting, the density of liquid water has a non-monotonic evolution close to the melting temperature with a maximal density at \blue{the temperature of maximum density $T_c$ equal to}  $3.98\,^\circ$C in fresh water, whereas the density increases linearly with solute concentration in the case of dissolution.  For ice melting in salt water, both temperature transport and salt transport contribute to determining melting velocity~\cite{Malyarenko2020,McCutchan2022,du2024physics}. This adds a further layer of complexity with opposing contribution of temperature and salt possibly leading  to double diffusive
convection flow~\cite{josberger1981laboratory}, controlling in some conditions the melting velocity~\cite{rosevear2021role,sweetman2024laboratory} and the shape of melting interfaces~\cite{xu2024buoyancy}. 
Finally, in contrast to the dissolution of salt and sugar, which are denser than the water bath, a buoy is not needed to ensure flotation, as ice floats on water (density $\rho_{ice}=916.7$\,kg\,m$^{-3} < \rho_{water}=999.8$\,kg\,m$^{-3}$ at the melting temperature $T_m=0\,^\circ$C~\cite{Handbook}). The contribution of melting to iceberg drift is currently unknown relative to other contributing factors that include ocean currents, wind, sea slope, surface waves, and Coriolis force~\cite{Mountain1980,Bigg1997,Wagner2017,Marchenko2019}.

Here, we investigate the kinematics of ice blocks which have asymmetric shapes while floating in a water bath and show that they can not only rotate but translate with significant speeds. We find a typical propulsion velocity of about 3\,mm\,s$^{-1}$ for triangular ice prisms with an inclined long side of approximately 20\,cm and width of approximately 10\,cm, floating in a water bath held at a temperature of about $T_b = 22\,^\circ$C, {comparable to those observed with dissolving bodies}. After presenting the experimental methods in Section \ref{Experimentalmethods}, we demonstrate and quantify the melting-driven propulsion mechanism in Section \ref{ExperimentalDemonstration}. We use shadowgraph imaging~\cite{settles2001schlieren,settles2017review} to simultaneously track the motion of the block and visualize the buoyancy convection flow. A phenomenological model relating the melting velocity to the terminal speed is developed in Section~\ref{sec:PropulsionModel} to explain the magnitude of the observed ice block translation velocities as a function of their size, inclination, and bath temperature. In the limit where the heat required to raise the temperature of the ice block to the melting temperature is relatively small, our model finds that although the latent heat plays an important role in the dynamics and determines the time over which the block melts, its actual magnitude does not significantly affect the propulsion speed. In Section~\ref{ParametricStudy}, we compare the measured values over a large range of parameters with the predictions of the model. 
Then, we demonstrate with experiments that the melting-driven propulsion mechanism extends to baths with ocean water salinity in Section \ref{saltwater}, and adapt our model to take into account the rising fresh water flow and the falling convection flow due to the cooling of the salt bath. We address the possible relevance of the melting-driven propulsion mechanism to icebergs in oceans in Section~\ref{Icebergs}, before summarizing our main results in Section~\ref{Discussion}. Although our results cannot be directly applied to icebergs under typical conditions, we argue that this effect may contribute to the motion of icebergs drifting in warmer subpolar regions. 

\begin{figure}[t]
    \centering
    \includegraphics[width=0.8\linewidth]{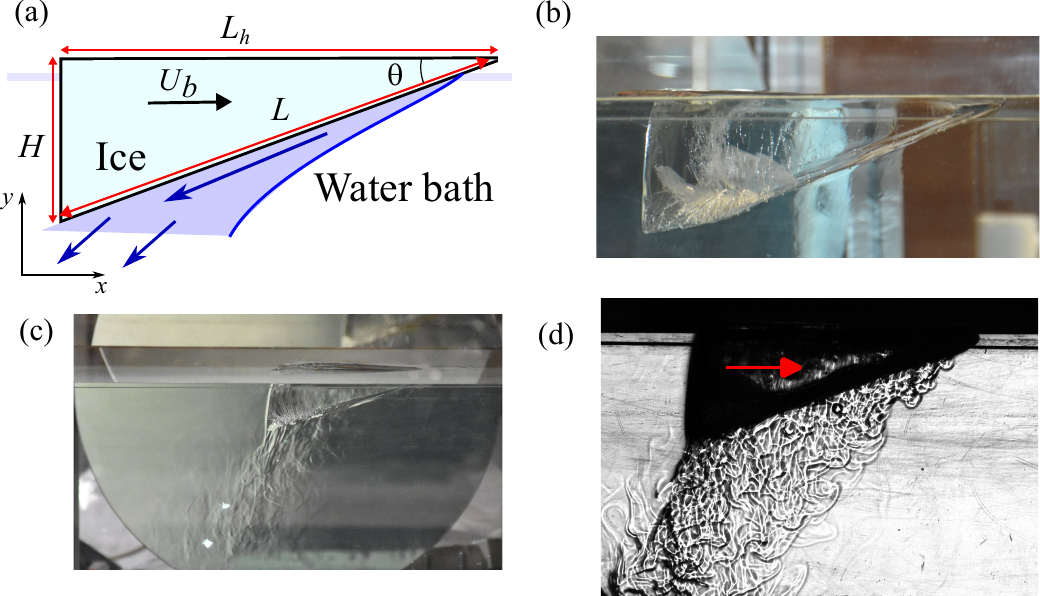}
    \caption{\blue{Principle of melting-driven propulsion experiments.} (a) A schematic of a melting ice block floating in a water bath. The block is a right-angled triangle prism. The hypotenuse has a length $L$ and is inclined at an angle $\theta$ relatively to the horizontal. The ice block propels along the horizontal coordinate $x$ with a velocity $\bm{U_b}$ in steady state.
    (b) A side view image of a right-angle ice block ($L_h = 163$\,mm, $W = 124$\,mm, $H = 65$\,mm, $\theta = 19.5^\circ$, $T_b=22\,^\circ$C).  
    (c) A profile view image of the floating ice block with a parabolic mirror placed behind the tank containing freshwater. The dense cold water leads to convection flow which is visible below the ice block. The mirror is then used for shadowgraph imaging. (d) A shadowgraph picture shows the ice block moving to the right with the convection directed toward the rear of the block (also see Movie~S1~\cite{sup-doc}). 
    }
    \label{fig:ice}    
\end{figure}

\newpage

\section{Experimental methods}
\label{Experimentalmethods}

We cast asymmetric ice blocks with a simple model geometry: a right-angled triangle prism. 
The ice block has a hypotenuse of length $L$ and is inclined at an angle $\theta$ with respect to the horizontal and is shown schematically in Fig.~\ref{fig:ice}(a). The ice blocks are cast in molds with horizontal length $L_h$, width $W$, and side $H$ filled with filtered demineralized water and freezing at $-15\,^\circ$C. Approximately one hundred blocks were cast with various sizes ranging over a few tens of centimeters. The ice blocks contain bubbles which typically results in about 10\% of air by volume, similar to icebergs~\cite{josberger1980effect}. A few clear rectangular ice blocks which are nearly transparent and defect-free were obtained from the Nice Company (https://www.thenicecompanyparis.com/fr), and cut into triangular prisms with a hot wire. Experiments performed with the clear ice blocks were found to give the same results as the ice blocks with trapped bubbles. The data sets corresponding to various blocks shapes and measurement protocols are listed in Appendix~\ref{Datasets}.  

Prior to the commencement of an experiment, an ice block is left to rest at room temperature of about $20\,^\circ$C for about ten minutes in order to avoid thermal shock when immersing it into the bath. During this time, the ice temperature approaches the melting temperature of ice.  Then, the block is carefully placed with its right angle on the top in a water bath with dimensions that are large compared to the block size. Flotation equilibrium corresponds to the vertical alignment of the gravity center and the center of the immersed part and is typically reached after few oscillations over a few seconds. A side view image of a floating ice block is shown in Fig.~\ref{fig:ice}(b). The actual hypotenuse inclination angle differs from $\theta$ by few degrees due to the buoyancy equilibrium of the asymmetric block (see Appendix~\ref{Sec:StabilityAnalysis}). Yet, we ignore this difference since it appears to have little effect on the measured trends.  
Ice blocks with $\theta > 39.6^\circ$ are gravitationally unstable, and we thus focus our study to ice blocks with $\theta$ below this angle.

The ice blocks and their motion in the bath are either observed with a camera from the side using a shadowgraph imaging~\cite{settles2001schlieren,settles2017review}, or from the top. To observe the dynamics with shadowgraphy, we use a glass tank with dimensions $116\times 46$ cm$^2$ filled up to $24$\,cm with tap water corresponding to a volume of about $128$\,liters. For most experiments the bath temperature $T_b$ is the ambient temperature. However, for a few experiments $T_b$ has been varied between $10.4$ and $30\,^\circ$C using a cooling / heating circulator \blue{before performing the measurements}. {A parabolic mirror with diameter $406$\,mm and focal length $1800$\,mm is placed on one side of the glass tank at a distance of about 100\,mm and oriented parallel to it, as depicted in Fig.~\ref{fig:ice}(c). A small Light Emitting Diode (LED) located at the focus of the parabolic mirror is used for illumination, and a digital camera is located \blue{also at the} focus of the mirror by the means of a semi-reflective plate. The resulting light beam with nearly parallel rays is refracted by the variations of optical indices due to temperature variations. The projected image captured by the digital camera integrates the density variations along the width of the tank.
 {For a bath of fresh water whose temperature is above maximum density of water at $T_c=3.98\,^\circ$C, the colder convection plumes sink and and refract ambient light. 
 The plumes appear more clearly in the image shown in Fig.~\ref{fig:ice}(d) generated with shadowgraphy. } {A small black circle marks the center of the mirror and is visible on some shadowgraph images and in the movies~\cite{sup-doc}. In these experiments, two 1~mm in diameter nylon wires are positioned just below the surface to guide the motion of the ice block, separated by a distance slightly greater than $W$. This helps maintain the distance between the block and the camera and limits rotation, facilitating observations. In the experiments where the ice blocks are viewed from the top, we use a glass tank with dimensions $90.5 \times 44.5$ cm$^2$, filled up to a height of at least $25$\,cm with filtered water corresponding to a volume of about $100$\,liters. The ice block is free to move in any horizontal direction and data is taken while the block is away from the tank sidewalls. This can be used to study the relative stability of the ice block trajectory.

\begin{figure}
    \centering
    \includegraphics[width=.85\textwidth]{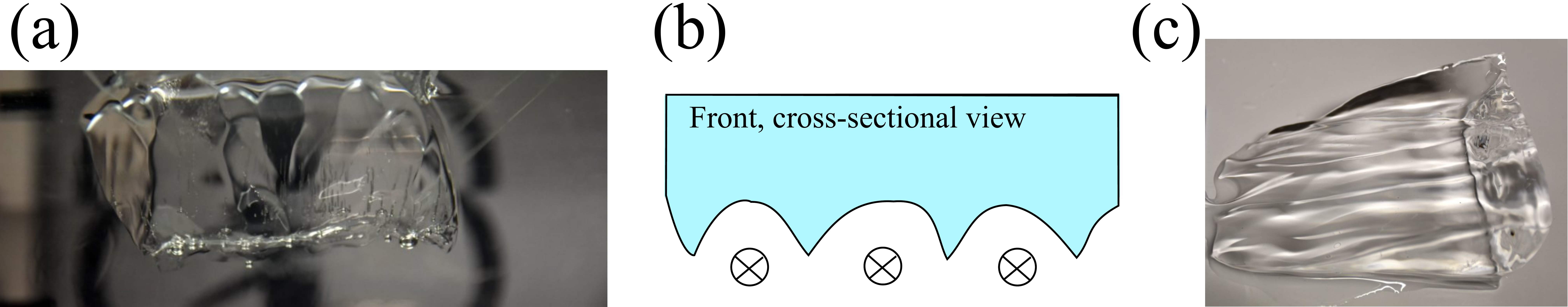}
      \caption{\blue{Observation of a melting pattern on the underside of ice blocks.} (a) A view of an ice block floating at the water surface while  moving  in the direction of the observer after being immersed in the bath for about 8\,minutes. A groove pattern generated by melting appears on the bottom inclined face. (b) A schematic cross section of the block illustrating the length-wise grooves. This melting pattern is carved by the backward descending flow. (c) An ice block  with bottom surface facing viewer after it is removed from the water bath following approximately 11\,minutes of immersion. Typical groove width and depth are about $20$\,mm and $5$\,mm, respectively. This example corresponds to Data Set \textbf{L}. ($L=176$ mm, $\theta=19.5^\circ$, $T_b=22.1\,^\circ$C).
      }  
            \label{FigGrooves}
\end{figure}

\newpage

\section{Experimental demonstration of melting-driven propulsion}
\label{ExperimentalDemonstration}

{Once the ice block is placed in the water bath, it begins to move after a transient period of a few tens of seconds, as a thermal convection flow develops below the melting block.} 
Figure~\ref{fig:ice}(d) and Movie~S1~\cite{sup-doc} show that this flow detaches and follows the inclined side of the block, creating a current from the front tip to the back. Consequently, by reaction, the ice block accelerates in the opposite direction. Over time, the ice block reaches a terminal velocity when the drag balances the propulsion force. We observe the emergence of concave grooves surrounded by crests along the inclined side. An image of the ice block and a schematic are shown {in Fig.~\ref{FigGrooves}.} 
This melting pattern is a generic feature of  ablation~\cite{Chaigne2023emergence} and is likely caused by the convection plumes advected by the mean current. Similar grooves have been also reported in simulations~\cite{couston2021topography} {as well as in experiments on the bottom side of horizontal ice cylinders melting in a warm, low-salinity bath~\cite{fukusako1994melting,yamada1997melting}}. Notably, these patterns do not appear to affect the robustness of the translation motion {in our experiments}. 

\begin{figure}[h]
    \centering
    \includegraphics[width=1\linewidth]{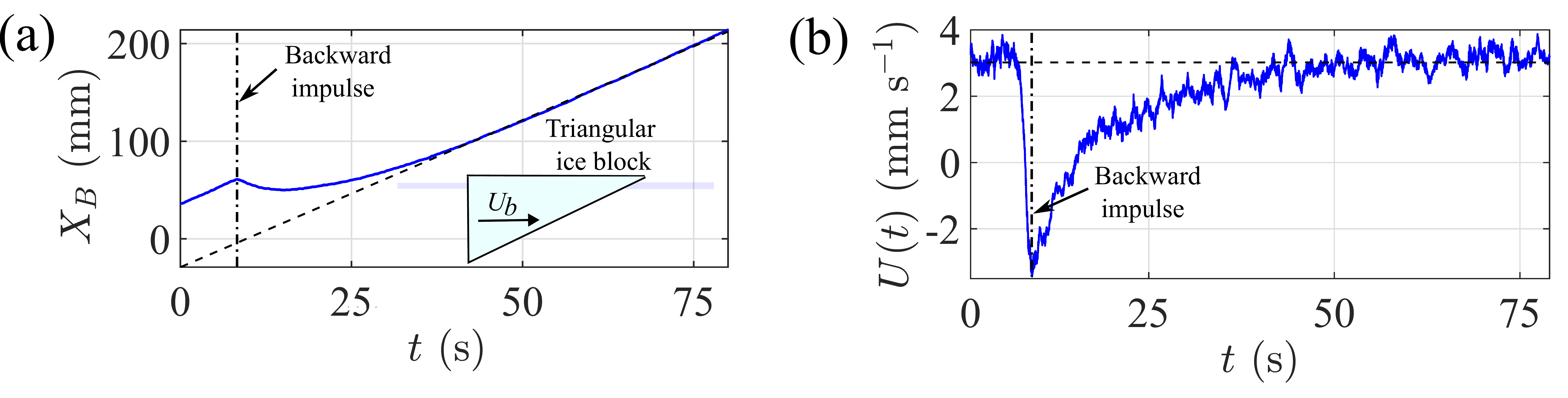}
    \caption{\blue{Demonstration of the self-propulsion for asymmetric melting ice blocks.} (a) Right-angle triangle horizontal ice block position $X_B$ as a function of time $t$ plotted along with a linear fit (dashed line) corresponding to a slope $U_b=3.02$\,mm\,s$^{-1}$. \blue{Data Set \textbf{G}. (b)} The velocity of the block $U(t)$ obtained from $X_B$. This experiment corresponds to the block depicted in Movie S2~\cite{sup-doc}, where the block is kicked manually in the direction opposite to the motion induced by propulsion at $t=8$\,s. After a transient, the block recovers the same velocity (dashed line).}
    \label{fig:FigExpdem}    
\end{figure}

To illustrate that the forward motion is quite robust, we manually kicked a block moving at constant velocity in the direction opposite to its motion at approximately 5\,s (see Movie~S2~\cite{sup-doc}). The measured ice block position $X_B$, obtained by processing the shadowgraph images, is plotted as a function of time $t$ in Fig.~\ref{fig:FigExpdem}(a).  The instantaneous velocity $U(t)$ is computed over a moving $1$\,s time interval and plotted in Fig.~\ref{fig:FigExpdem}(b). The terminal velocity $U_b$ is obtained by fitting a line to Fig.~\ref{fig:FigExpdem}(a) once steady motion is resumed. We plot a dashed horizontal time in Fig.~\ref{fig:FigExpdem}(b) and observe that the block 
accelerates and reaches the pre-kick velocity $U_b \approx 3.02$\,mm\,s$^{-1}$ in about 30\,s. 
By contrast, symmetric rectangular ice blocks do not show net translation motion (see Movie~S3~\cite{sup-doc}). {We plot $X_B(t)$ and $U(t)$ for this symmetric ice block in Fig.~\ref{fig:MES40807}(a), and Fig.~\ref{fig:MES40807}(b), respectively. The time scales for motion are significantly larger compared with those observed in the asymmetric examples. The ice block motion shows no clear direction, and capsizing events~\cite{JohnsonPRF2025} occur suddenly and modify the convection flow that can change the direction of the ice block drift. 

\begin{figure}
    \centering
    \includegraphics[width=0.85\textwidth]{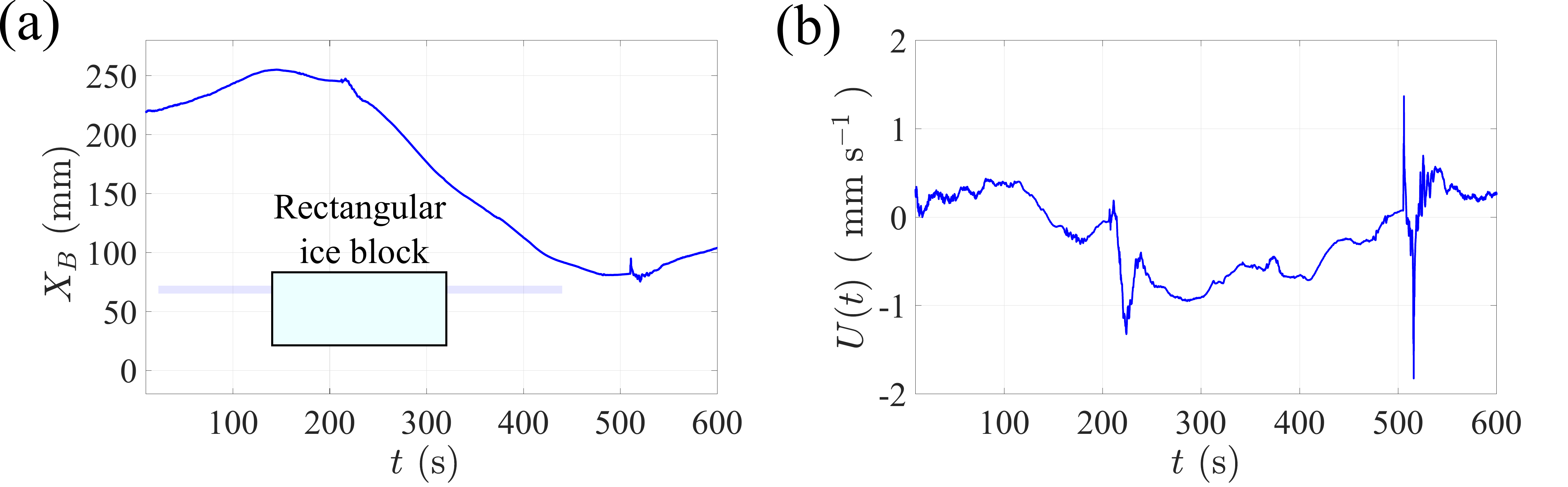}
 \caption{\blue{Erratic motion for symmetric melting ice blocks.} (a) For a symmetric rectangular ice block position $X_B$ shows the slow drift of the block over time (also see Movie S3~\cite{sup-doc}). The block dimensions: $100\times 40 \times 40$ mm$^3$, $\theta=0^\circ$, $T_b=20.1\,^\circ$C. \blue{(b)} The corresponding velocity (averaged over $10$~s) shows sudden changes in velocity due to capsizing events at $t=210$\,s and $t=506$\,s.}
    \label{fig:MES40807}
\end{figure}

{We also perform experiments with ice blocks which are unconstrained laterally and we observe their motion in the tank from above.} While the sides of the block melt on average at a rate of a few tens of microns per second, we observe that the block inclination is approximately conserved, at least during the first ten minutes or so, even as the ice block shrinks and the edges round over time. However, we find that the movement is not exactly rectilinear, but can show rotation in addition to translation. This rotation may be due to lateral asymmetries in the ice block, and/or caused by the destabilization of the falling convection flow into a vortex as reported with spinning ice disks~\cite{Dorbolo2016,schellenberg2023rotation,li2025hydrodynamic}. The rotational motion becomes increasingly important over time, as the ice block decreases in size and its shape deviates significantly from being prismatic as it completely melts. For this reason, systematic data is taken during the first ten minutes while the overall ice block shape appears similar as the one initially cast, barring the grooves. In these conditions, the dimensions of the blocks considered are determined at the start of the run. The different protocols with various ice blocks shapes are summarized in Appendix~\ref{Datasets}.

\blue{The measured terminal velocities} $U_b$ show more significant experimental variability compared with dissolution propelled bodies~\cite{Chaigne2023} that can be attributed to several factors. The density contrast driving the convection flow is stronger by two orders of magnitude in the case of sugar or salt dissolution compared with the density contrast observed in ice blocks when $T_b = 20\,^\circ$C. \blue{Consequently, the ice block motion is more sensitive to external perturbations and experimental imperfections.} 
The typical scale of the plumes about $3\delta_i \approx 5$ mm as discussed in Section~\ref{ModelMelting} is larger compared to the block size leading to greater fluctuations in their trajectories. The shapes of the ice blocks are also less ideal and show around 5\% variations. The ice blocks also intermittently release trapped bubbles during melting that further perturb transiently the flow below the ice block. Recent experiments~\cite{wengrove2023melting} report an enhanced melting rate for vertical ice walls, due to the additional buoyancy associated with the released bubbles. However, in our experiments, we do not find significant differences on average for $U_b$ between the standard ice blocks and those made of clear ice without bubbles. 
Notwithstanding the fluctuations, we observe a robust directed motion with velocity $U_b \approx 3$\,mm\,s$^{-1}$ for an ice block with $L= 20$\,cm melting in a bath at 22$\,^\circ$C which can be easily noted with the naked eye. The corresponding Reynolds number $Re= \frac{L\,U_b}{\nu}$ is of order $600$, with the kinematic viscosity of water $\nu \approx 10^{-6}$ m$^2$ s$^{-1}$. Thus, the fluid flow is in the inertial regime, and we propose a phenomenological model based on momentum balance inspired by Chaigne, {\it et al.}~\cite{Chaigne2023}.}

\section{Propulsion model for inclined ice blocks melting in freshwater} 
\label{sec:PropulsionModel}

\begin{figure}[h]
    \centering
    \includegraphics[width=.55\linewidth]{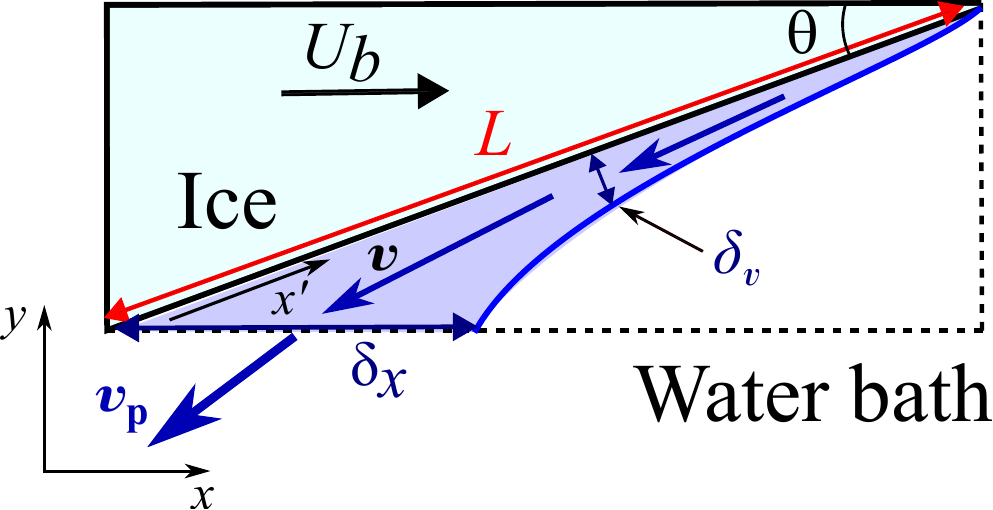}
    \caption{\blue{Schematic of the melting driven propulsion mechanism. A} time-averaged current due to the thermal convection flow  with velocity $\bm{v}$ develops below the ice block in a fluid layer with thickness $\delta_v$ which increases with distance from the front tip. \blue{$x'$ is the coordinate along the inclined wall from the rear to the tip of the ice block.} The current leaves the control volume indicated by the dashed line below the ice block, over a length $\delta_x$ with a velocity $\bm{v_p}$, generating a horizontal thrust in the opposite direction which is balanced by fluid drag in the stationary regime. Consequently, the ice block moves in the $x$ direction with a steady velocity $\bm{U_b}$.}
    \label{MeltingSchemaA}
\end{figure}

\begin{figure}[h]
    \centering
    \includegraphics[width=.6\textwidth]{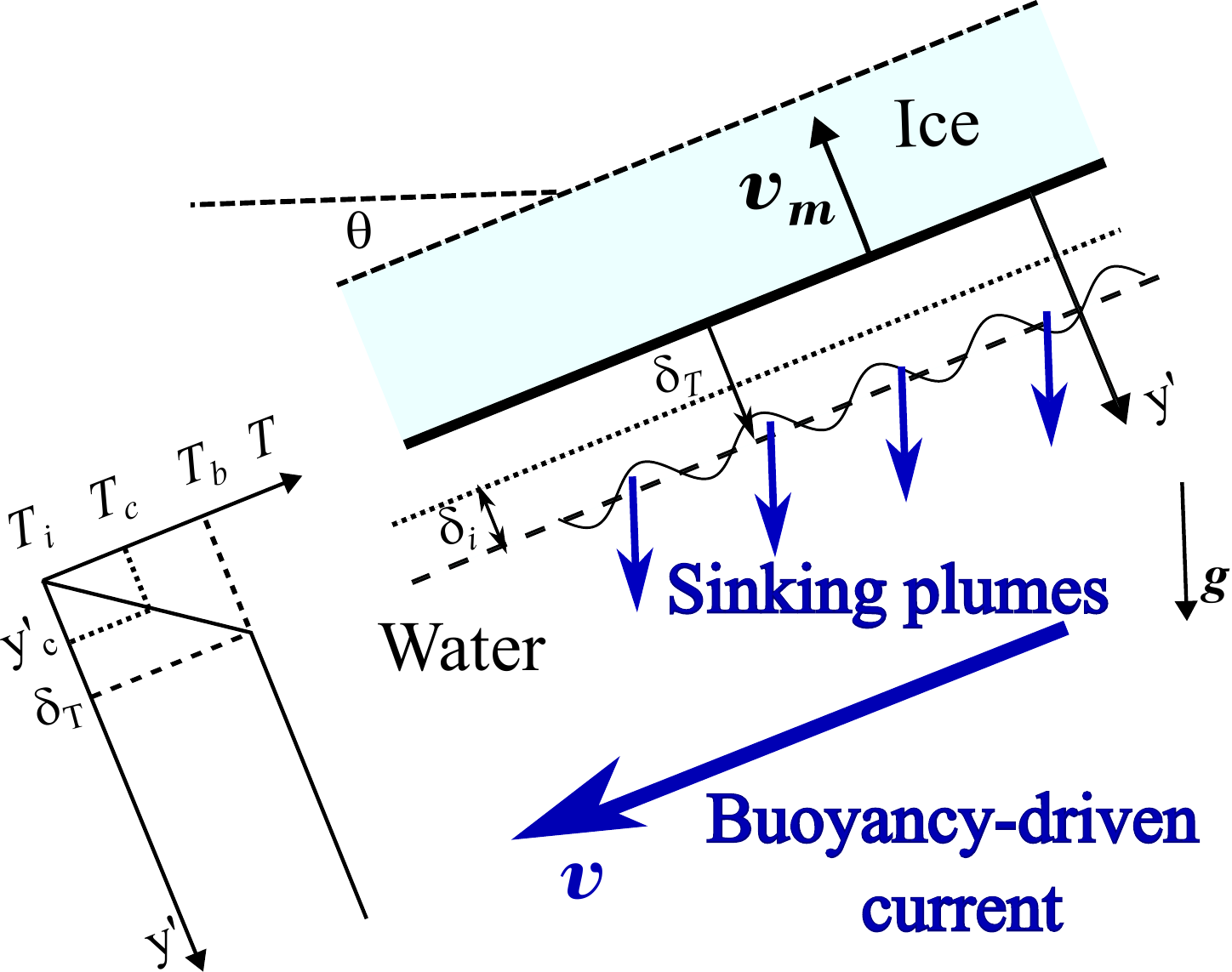}
    \caption{A schematic section of the melting ice-water interface and the current. The velocity of the ice/water interface during melting $\bm{v_m}=-v_m\, \bm{e_{y'}}$, defines the melting velocity $v_m$. The graph on the left depicts the temperature evolution as a function of the distance to the block $y'$. The thermal boundary layer is assumed to have a constant thickness $\delta_T$ under steady melting conditions. For simplification, we assume that the increase of the temperature between the ice temperature $T_i \equiv T_m = 0\,^\circ$C and the bath temperature $T_b$ is linear. \blue{Because the water density is maximal at \blue{the temperature of maximum density $T_c$}, the domain for which $T_m < T< T_c$ is buoyancy stable, whereas the domain for which $T_c < T< T_b$ is instable. We consider the instability of this layer of thickness $\delta_i$ , which emits sinking plumes. These plumes feed a gravity current that moves along the inclined surface with velocity $\bm{v}$. By linearity of the temperature profile, the estimation of $\delta_T$ is deduced from the one of $\delta_i$.}}
    \label{SchemaMeltingB}
\end{figure}

We develop here a two-dimensional (2D) model which explains the propulsion mechanism, provides a quantitative analysis of our observations, and predicts the magnitude of the terminal velocity $U_b$ following the approach developed for dissolving blocks by Chaigne, {\it el al.}~\cite{Chaigne2023}. We consider the central vertical plane and the surface below the ice block as shown in Fig.~\ref{MeltingSchemaA}. Because significant water flow is not observed near the vertical back surface in the shadowgraph images, we ignore the back surface, and also neglect the lateral flat sides, which by symmetry do not generate a net propulsion contribution. The heat extracted from the bath to melt the ice leads to a cooling of the bath water near the ice block. For $T_b > T_c$ in freshwater, the cooled water is denser and drives the convection current downwards and to the back with an average velocity $\bm{v}$. Similar to dissolving blocks~\cite{Chaigne2023}, a propulsion force $F_c$ is generated by a time-averaged density driven convection current that escapes a control volume below the block with a velocity $\bm{v_p}$.
Although models of melting-driven propulsion and of dissolution-driven propulsion are similar, we identify significant differences. The melting velocity is computed using the Stefan condition~\cite{AlexiadesSolomonBook} expressing energy conservation instead of the dissolution boundary condition expressing solute conservation~\cite{carpy2024fingerprints}. The density changes are due to the temperature variations instead of solute concentrations. Due to the non-linear equation of state of liquid water, the density has a non-monotonic variation with temperature instead of a linear variation. In both cases, the thickness of the unstable boundary layer is evaluated using a criterion for turbulent convection by computing either the thermal Rayleigh number for melting or the solutal Rayleigh number for dissolution.

\subsection{Melting driven by thermal convection}
\label{ModelMelting}
{We evaluate the melting velocity $v_m$ of a floating inclined ice block valid when $T_b > T_c = 3.98\,^\circ$C, the temperature where freshwater density is maximum. The velocity of the receding ice/water interface during melting is directed locally perpendicular to this interface and its magnitude defines $v_m$ (see Fig.~\ref{SchemaMeltingB}). We consider the melting as being driven by thermal convection because the block is initially stationary, and the water below the block is unstable due to gravity as it cools.} Keitzl, {\it et al.}~\cite{keitzl2016} investigated experimentally and numerically the melting of ice in fresh water for a \blue{horizontal ice lid immersed in a water bath}. These authors also derived  a scaling law for the melting velocity driven by thermal convection. However, their model is arbitrarily calibrated using  numerical simulations. Here, we propose a model to estimate the melting velocity $v_m$ when the interface is inclined, and shown to be consistent with that by Keitzl, {\it et al.}
~\cite{keitzl2016}. 

We consider an ice block with a volume which is much smaller than the water bath volume, and assume that the bath temperature $T_b$ is constant, sufficiently far from the block. The ice block surface temperature $T_i$ is assumed to be at the ice melting temperature $T_{m}=0\,^\circ$C. 
The latent heat of ice melting \blue{is} $ \mathcal{L} = 333.5$~kJ\,kg$^{-1}$, and the heat capacity of ice is $C_{p,i}=2.11$~kJ\,kg$^{-1}$\,K$^{-1}$. The energy required to heat the ice block to $T_i$ 
for an initial ice block temperature of $-15\,^\circ$C can be thus considered to be negligible, because  $\mathcal{L}/C_{p,i} \gg 15$\,\blue{K}.  Further, the ice blocks are allowed to rest in our experiments for at least $\tau_{rest}=10$~minutes after removal from the freezer. As the ice diffusivity is $\kappa_{ice}=1.11\times 10^{-6}$~m$^2$\,s$^{-1}$, the temperature typically diffuses over a distance $\sqrt{\kappa_{ice} \, \tau_{rest}} \approx 26$\,mm. Therefore, {we neglect the heat transport inside the block for ice blocks that are a few tens of centimeters in size, and assume that the inner temperature of the ice blocks near the melting interface is $T_m$}.

Water density has a non-monotonic dependence on temperature $T$, and can be modeled approximately by a quadratic polynomial function~\cite{keitzl2016}:
\begin{equation}
\rho(T)=\rho_c\,[1-\beta \,(T-T_c)^2] \, , 
\label{rhoeau}
\end{equation}
 where $\rho_c=999.96$ kg m$^{-3}$ is the maximal water density at  $T_c=3.98\,^\circ$C, and coefficient $\beta=7 \times 10^{-6}$ K$^{-2}$. Thus, $\rho$ of water near the ice block is greater than that of the warmer bath water when $T_b > T_c$,  and this density difference drives a convection flow. The temperature dependence of $\rho$ is more accurately described by Bigg's empirical relation~\cite{bigg1967density,sharqawy2010thermophysical}. By fitting $\rho(T)$ given by Bigg's relation, the coefficient $\beta=6.5 \times 10^{-6}$ K$^{-2}$ in Eq.~(\ref{rhoeau}) is obtained, and used henceforth.  
Under steady state conditions, we assume turbulent thermal convection, \textit{i.e.} on average the temperature change is localized to a thin thermal boundary layer of thickness $\delta_T$, where the temperature increases from $T_i$ to $T_b$ as illustrated in Fig.~\ref{SchemaMeltingB}. To simplify the modeling, we assume a linear temperature profile. The heat transport is diffusive in the boundary layer and convective outside. In the turbulent region below, the sinking plumes and upwelling flow transfer heat \blue{more efficiently than thermal diffusion}. Due to the inclination of the block, the convection flow self-organizes into a current with characteristic velocity $\bm{v}$ 
on the scale of the ice block, which escapes the control volume with a velocity $\bm{v_p}$ which is directed below and behind the block 
(see Fig.~\ref{MeltingSchemaA}). This directed flow is the origin of the propulsion mechanism.

We consider \blue{the case where the melting velocity $v_m$ of the solid-liquid interface is controlled by the convection flow}.  We denote $y'$ as the coordinate normal to the melting interface and $y'_c$ as the distance to where $T =T_c$. Because $\rho$ is below $\rho_c$ where $T_i<T<T_c$ over $0<y'<y'_c$, the fluid layer is stable \blue{relative to a density inversion instabilities due to gravity, leading to detached flows as in the Rayleigh-B\'enard or Rayleigh-Taylor instabilities}. In contrast, the domain $y'_c<y'<\delta_T$ is denser than the bath at the density $\rho(T_b)$ and can be susceptible to \blue{such convection instabilities}. Assuming a linear temperature profile, we have $y'_c=\delta_T\,(T_c-T_i)/(T_b-T_i)$ and consequently the  thickness of the unstable layer is $\delta_i=\delta_T-y'_c=\delta_T\,(T_b-T_c)/(T_b-T_i)$. We evaluate the dimensionless density contrast by dividing $\Delta \rho$ the density difference between the fluid at $T_c$ and the fluid at $T_b$, by the average density between these two values:
\begin{equation}
\dfrac{\Delta \rho}{\rho}= 2 \left(\dfrac{\rho_c-\rho_b}{\rho_c+\rho_b} \right)\,,
\label{Deltarho}
\end{equation}
where $\rho_b=\rho (T_b)$ is computed using Eq.~\ref{rhoeau}. 
\blue{Large enough values of $\Delta \rho$ may trigger Rayleigh-B\'enard-like instabilities.} According to previous studies in geometries with semi-infinite extent under steady state conditions~\citep{Malkus1954,Sullivan96,Philippi2019}, and {as discussed in a lecture by Linden~\cite{LindenConvection},} the thickness of the boundary layer remains on average close to the critical value at the onset of the Rayleigh-B\'enard instability to first approximation. From the definition of the Rayleigh number $Ra$ and assuming the critical value of $Ra=Ra_c$, we have
\begin{equation}
Ra_c=\dfrac{\Delta \rho}{\rho} \dfrac{\delta_i^3\,g\, \cos \theta }{\kappa \, \nu},
\label{eq:Rac}
\end{equation}
where $g \,\cos \theta$ is the projection of the gravitation acceleration perpendicular to the ice block surface and $\kappa$ is the thermal diffusivity of water. To obtain an estimate, we use the parameter values from Keitzl,~\textit{et al.}~\cite{keitzl2016}, where their temperature dependence is neglected. We take the value of $\kappa=1.33\times 10^{-7}$\,m$^2$ s$^{-1}$ at $T=0\,^\circ$C, because between $T=0\,^\circ$C and $T=30\,^\circ$C, $\kappa$ changes less than $10\%$~\cite{Handbook}. However, the decrease in $\nu$ with temperature is more significant because $\nu=1.79\times 10^{-6}$ m$^2$ s$^{-1}$ at $T=0\,^\circ$C, $\nu=1.58\times 10^{-6}$ m$^2$ s$^{-1}$ at $T=4\,^\circ$C and $\nu=1.00 \times 10^{-6}$ m$^2$ s$^{-1}$ at $T=20\,^\circ$C~\citep{Handbook}. Accordingly, the thermal Prandtl number in water  $Pr=\nu/\kappa$ is about $13.5$ at $T=0\,^\circ$C, $11.9$  at $T=4\,^\circ$C, and $7.5$ at $20\,^\circ$C. To simplify, we choose the average value of $\nu$ between $T_c$ and $T_b$. 

{Because the unstable layer of thickness $\delta_i$ is taken between the two stress-free interfaces,} $Ra_c = \frac{27}{4}\,\pi^4 \approx 658$~\citep{Chandrasekhar}. Consequently, we find $\delta_T$ as:
\begin{equation}
\delta_T=\dfrac{T_b-T_i}{T_b-T_c}\,\delta_i\, , \quad \mathrm{with} \quad \delta_i=\left( \dfrac{Ra_c\, \kappa\,\nu}{g\,\cos \theta}\right)^{1/3} \, \left( \dfrac{\Delta \rho}{\rho} \right)^{-1/3} \, ,
\label{deltaT}
\end{equation}
after rearranging Eq.~(\ref{eq:Rac}). For $T_b=20\,^\circ$C and $\theta=26.5^\circ$, $\nu=(\nu(T_c)+\nu(T_b))/2=1.29\times 10^{-6}$ m$^2$ s$^{-1}$ and $\Delta \rho / \rho \approx \beta\,(T_b-T_c)^2\approx 1.80 \times 10^{-3}$, we find $\delta_i \approx 1.93$\,mm and $\delta_T \approx 2.40$\,mm. 
The experimental and numerical study by Du, \textit{et al.}~\cite{du2023sea} for freezing of salt water provide correlation laws for $\nu$ and $\kappa$. With these more accurate values of the fluid properties, we find $\delta_T \approx 2.44$~mm, which is close to $\delta_T \approx 2.40$\,mm obtained using Eq.~(\ref{deltaT}). 

The wavelength at the marginal instability is approximately {3$\delta_i$}~\cite{Chandrasekhar}, which gives a typical plume size of $5.8$\,mm. {This spacing between plumes is not visible in shadowgraph images of the inclined blocks because the flow shows a complex pattern of entangled plumes. However, it can be seen more clearly for horizontal blocks (see Movie~S3~\cite{sup-doc}). In this case, we find an average distance between plumes in steady convection regime of about $5$~mm, which is close to the wavelength expected for the marginal instability. This characteristic length for melting plumes is significantly larger than for the plumes} caused by dissolution of salt or sugar in water, which are approximately 0.33\,mm and 1\,mm, respectively~\cite{Philippi2019,Cohen2020,Chaigne2023}.

Assuming negligible heat flux in the ice, $v_m$ is obtained from the Stefan condition~\cite{AlexiadesSolomonBook} at the melting interface:
 \begin{equation}
v_m=  \dfrac{\rho_i}{\rho_{ice}}\,\dfrac{ C_p\,\kappa}{\mathcal{L}}\,\dfrac{\partial T}{\partial y'}\Big{\vert}_{y'=0} \,,
 \label{Stefancond}
\end{equation}
where $\rho_i=\rho (T_i)$ and $C_p \approx 4.2$~kJ~kg$^{-1}$~K$^{-1}$ is the heat capacity of liquid water at the melting temperature $T_i$.  

Assuming a linear temperature profile across $\delta_T$, 
$ \dfrac{\partial T}{\partial y'}\Big{\vert}_{y'=0} = \dfrac{(T_b-T_i)}{\delta_T}$, and 
using Eq.~(\ref{deltaT}), 
\begin{equation}
    v_m  =  \dfrac{\rho_i}{\rho_{ice}}\,\dfrac{\Gamma C_p\,\kappa}{\mathcal{L}}\,\dfrac{T_b-T_i}{\delta_T},
\end{equation}
where $\Gamma$ is a dimensionless constant introduced to take into account the deviation from a linear temperature profile in experiments. The Stefan number ${St}_b$ gives a measure of the energy required to cool the water bath compared to the latent energy, and is given by ${St}_b=\dfrac{C_p\,(T_b-T_c)}{\mathcal{L}}$. After replacing $\delta_T$ using Eq.~(\ref{deltaT}), we find  
\begin{equation}
    v_m  =  \dfrac{\rho_i}{\rho_{ice}}\,\Gamma \,{St}_b\,(Ra_c\,Pr)^{-1/3}\,\left( \dfrac{\Delta \rho}{\rho} \right)^{1/3} \, (\kappa \,g \,  \cos \theta )^{1/3} \, .
\label{vmelting} 
\end{equation}

To find the fitting constant $\Gamma$, a set of measurements of $v_m$ were performed with inclined blocks. Clear ice blocks with dimensions $100\times 40 \times 100$ mm$^3$ were fixed at a prescribed inclination over the range of $\theta$ where the ice blocks were found to be stable. The displacement of the bottom interface while subjected to a detached thermal convection flow was monitored using shadowgraph imaging and backlight imaging. The \blue{measured melting velocities} $v_m$ with these two methods are shown in Fig.~\ref{vmelting}(a). While the measurement scatter is considerable, the two methods yield $v_m$ which are consistent with each other. Then, we estimate $\Gamma=2.2$ by fitting Eq.~(\ref{vmelting}) to the data in Fig.~\ref{fig:Meltingrate}(a). Because the angle dependence is $(\cos\theta)^{1/3}$, the dependence on angle is small, and within measurement scatter.

\begin{figure}
    \centering
        \includegraphics[height=0.35\textwidth]{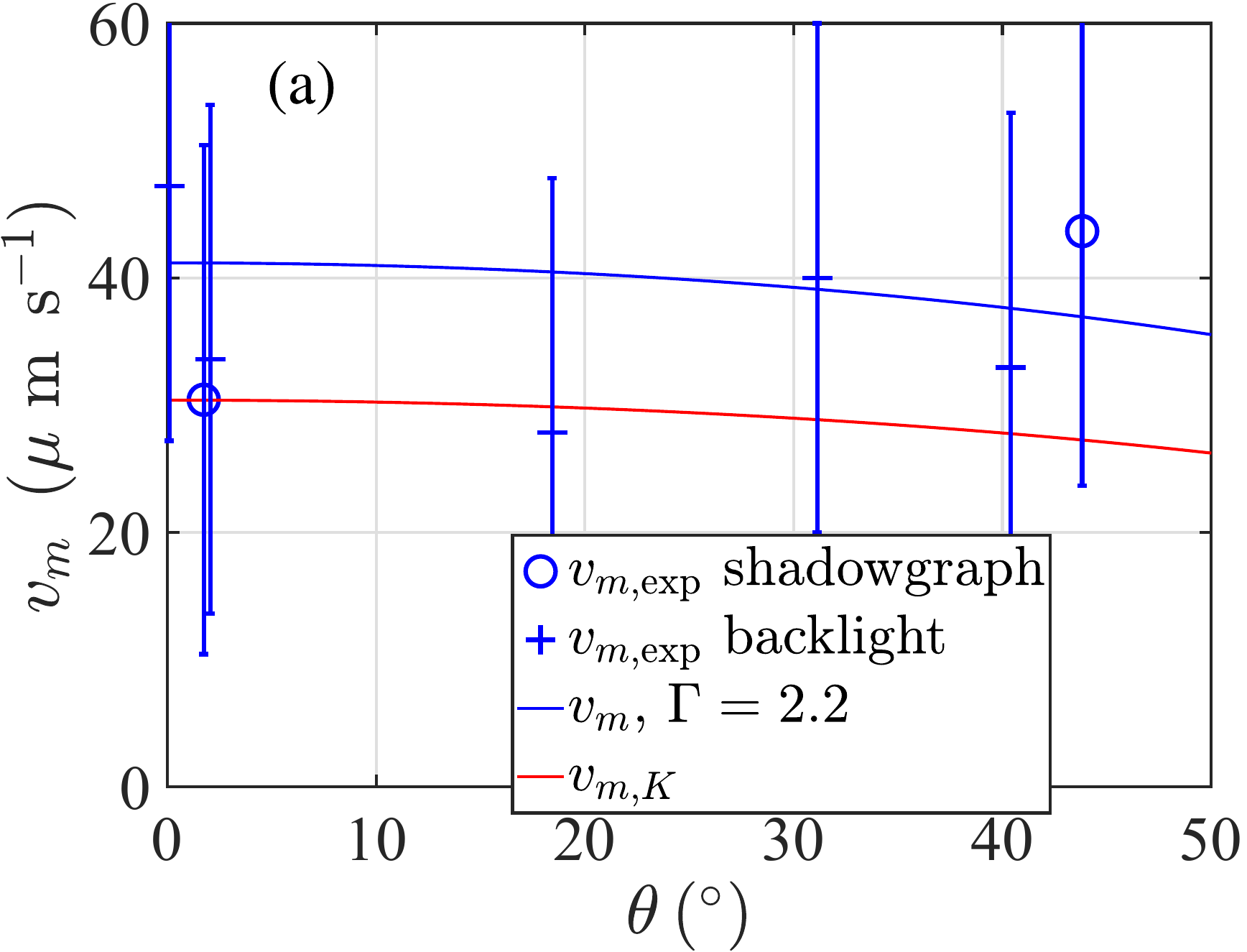} \hfill
      \includegraphics[height=.35\textwidth]{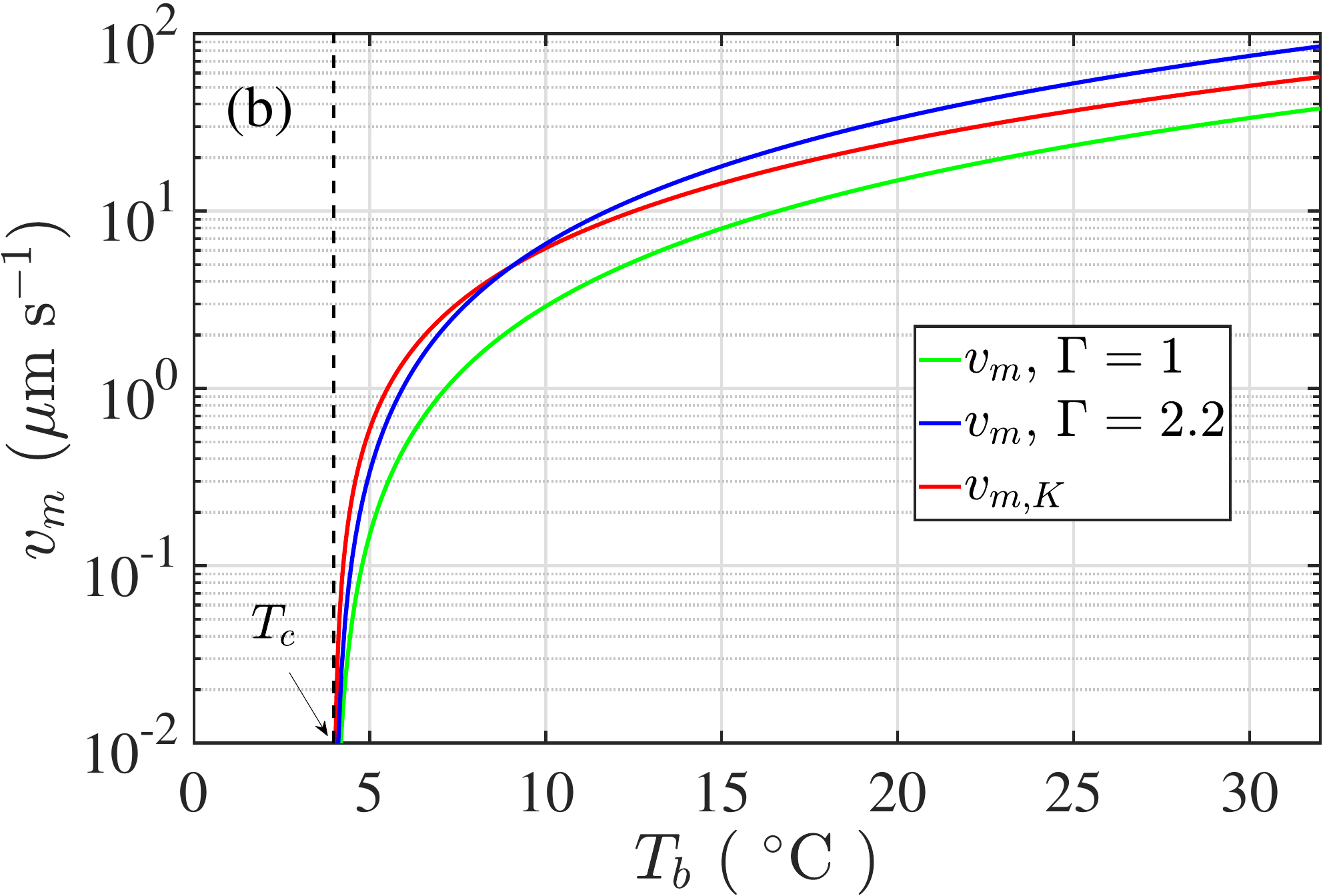}
    \caption{\blue{Measurements and theoretical prediction of melting velocity of ice in fresh water.} (a) 
    The experimental measurements of $v_m$ at the bottom surface of an ice block as a function of its inclination in fresh water. The blocks are held at rest, $T_b=22\,\,^\circ$C and their dimensions correspond to Data Set {\bf F} (clear ice). Two illumination methods are used, shadowgraph and backlight. \blue{The errorbars correspond here to a measurement uncertainty of $\pm$ 20 $\mu$m s$^{-1}$.} These measurements enable us to calibrate the ice melting model in fresh water and set $\Gamma = 2.2$ in Eq.~(\ref{vmelting}). The theoretical estimate of Keitzl, \textit{et al.}~\cite{keitzl2016} after including an inclination dependency according to Eq.~(\ref{eq:vmk}) predicts relatively well the measurements. (b) The calculated melting velocity of the ice block in freshwater with $\theta=24^\circ$ as a function of \blue{the bath temperature} $T_b$ using Eq.~(\ref{vmelting}) for $\Gamma=1$ and $\Gamma = 2.2$. The theoretical estimate of Keitzl, \textit{et al.}~\cite{keitzl2016} $v_{m,K}$ according to Eq.~(\ref{eq:vmk}) is also plotted. Both models are valid for $T>T_c$. \blue{For $T<T_c$, the melting velocity results from heat transport by thermal diffusion only. This case is not addressed by our melting model driven by thermal convection.} } 
    \label{fig:Meltingrate}
\end{figure}

The melting velocity estimated by Keitzl, \textit{et al}.~\cite{keitzl2016} \blue{$v_{m,K}$}, after multiplying gravitational acceleration $g$ by a factor $\cos \theta$ to include the effect of the block inclination, can be written as,
\begin{eqnarray}
    v_{m, K} &=& v_{m,K0} \, \left[ \dfrac{(T_b-T_c)^2}{(T_b-T_m)\,(T_c-T_m)} \right]^{2/3} \, \left( \dfrac{(T_b-T_m)+(T_c-T_m)}{T_c-T_m} \right), \,\,\mathrm{where} \label{eq:vmk} \\  \,\, v_{m,K0} & = &  \dfrac{\rho_b}{\rho_{ice}} \,\dfrac{C_p\,(T_c-T_m)}{\mathcal{L}} \,  \left( \dfrac{Ra_c\,Pr}{4} \right)^{-1/3}\,\left(\dfrac{\beta\,T_c^2}{\rho_b}\right)^{1/3} \left(\kappa\,g \cos\theta  \right)^{1/3} \, , \nonumber
\end{eqnarray}
and $\beta\, T_c$ is a density contrast according to Eq.~(\ref{rhoeau}) between the maximal water density and the water density at $T_m=0\,^\circ$C. They considered $Pr=10$ and $Ra_c=1000$, comparable to the values in our study, and calibrated the temperature dependency using numerical simulations of the temperature profile at the melting interface~\cite{keitzl2016} while testing the predicted values only over temperatures ranging from $4.5\,^\circ$C to $14.8\,^\circ$C with experiments. 
Plotting $v_{m, K}$ in Fig.~\ref{fig:Meltingrate}(a), we observe that it is somewhat lower compared with our experimental data, but reasonable considering the apparent experimental fluctuations. We also plot $v_{m, K}$ using Eq.~(\ref{eq:vmk}), and our calculated $v_m$ using Eq.~(\ref{vmelting}) with fit $\Gamma=2.2$, {as well as with \blue{\sout{fit}} $\Gamma=1$} 
in Fig.~\ref{vmelting}(b) as a function of \blue{the bath temperature} $T_b$ for $T_b > T_c$, where both models are valid. The two models provide similar predictions of melting velocity and the same scaling at large $T_b$. 

{Further, our calculation of $v_m$ (Eq.~\ref{vmelting}) gives the same scaling as previously found by Kerr in Appendix~B of Ref.~\cite{KerrJFM1994}, but without taking into account the nonlinear evolution of water density with temperature. The effect of this nonlinear evolution has been investigated in numerical studies of Rayleigh-B\'enard instability in presence of a upper stable stratified layer above a unstable region~\cite{Toppaladoddi2018,olsthoorn2021cooling,couston2021conv}, but without phase change. Under standard conditions for ice melting, the dynamics of the upper stable layer may be decoupled from the thermal convection flow below~\cite{couston2021conv}.  As in Kerr~\cite{KerrJFM1994} and Keitzl~\cite{keitzl2016}, we neglect the contribution of the melt fluid in the calculation of the melting velocity. The layer of melt fluid at $T_m$ thickens the stable stratified layer, insulating the ice block, and reducing the melting velocity. Using numerical simulations, Keitzl \textit{et al.}~\cite{keitzl2016} estimate a decrease of less than $10$\% of the meting rate for bath temperature smaller than $20\,^\circ$C.}
Finally, after non-dimensionalizing Eq.~(\ref{vmelting}), and using a characteristic length scale $L^\star$ large compared to $\delta_T$, it can be shown that the dimensionless thermal flux, the Nusselt number $Nu$, is proportional to the Rayleigh number to the power $1/3$. In the general context of thermal convection, this scaling law corresponds to the regime where the heat flux is controlled by the thermal boundary layer~\cite{Malkus1954,castaing1989scaling,ahlers2009heat}. {The scaling $Nu \propto Ra^{1/3}$ has been also reported experimentally for the melting of ice cylinders about ten centimeter of diameters in fresh and saline water~\cite{yamada1997melting,bellincioni2024melting}}.

\subsection{Buoyancy current and terminal velocity}
We evaluate the magnitude of the time average current below the block at the point where it exits the control volume with a velocity $\bm{v_p}$ 
(see Fig.~\ref{MeltingSchemaA}). 
Because the heat extracted from the bath to melt the ice block cools the fluid as it moves from the front to the back, the temperature of the fluid $\hat{T}_b$ where it exits the control volume is lower than \blue{the bath temperature} $T_b$, and the corresponding density $\hat{\rho}_b$ is greater than $\rho_b$. 
The resulting density difference accelerates the current under the action of gravity.
To determine the density increase $\hat{\rho}_b - \rho_b$, we write the energy balance in a fluid layer of thickness $\delta_v$ moving parallel the interface with velocity $\bm{v}$ below the melting block. 
Noting that the energy flux is controlled by $v_m$ and that  the fluid outside the thin thermal boundary layer is well stirred {and assuming $v \approx v_p$}, we have to first order, 
 \begin{equation}
\rho_b\,v_p\,\delta_v\,C_p\,(\hat{T}_b-{T}_b)=-\rho_{ice}\,\mathcal{L}\,L\,v_m 
\, ,
  \label{deltaTcooling}
 \end{equation} 
Thus, the temperature below the block becomes,
 \begin{equation}
\hat{T}_b-{T}_b=-\dfrac{\rho_{ice}}{\rho_b}\,\dfrac{
v_m}{v_p}\,\dfrac{L}{\delta_v}\,\dfrac{\mathcal{L}}{C_p} \,.
\label{Eq:Tbhat}
 \end{equation} 
Then, using Eq.~(\ref{rhoeau}) to evaluate the corresponding density change, we have,
\begin{equation}
 \hat{\rho}_b-\rho_b =- \beta \, \rho_c\, (\hat{T}_b- T_b)\,\left[ (\hat{T}_b - T_b) + 2 (T_b-T_c) \right].
 \label{deltarhocoolinggen}
\end{equation}
 For small temperature difference compared to $T_b-T_c$, $ \hat{\rho}_b-\rho_b \approx - 2 \beta \rho_c (\hat{T}_b- T_b)\,(T_b-T_c)$, and we obtain, 
 \begin{equation}
 \hat{\rho}_b-\rho_b=\dfrac{\rho_{ice}}{\rho_b}\dfrac{2\beta \rho_c (T_b-T_c)\,\mathcal{L}}{C_p} \, \dfrac{{L}\,
v_m}{\delta_v\,v_p}.
  \label{deltarhocooling}
 \end{equation} 
In practice, this approximation for evaluating $\hat{\rho}_b$ is well justified. As in Ref.~\cite{Chaigne2023}, the velocity of the gravity driven current is then set by balancing gravity force due to the density increase given by Eq.~(\ref{deltarhocooling}) with inertial drag with a dimensionless friction coefficient $f_D$,
\begin{equation}
( \hat{\rho}_b-\rho_b)\,g\,{L}\,\sin \theta=f_D\,\rho_b \, \dfrac{{L}}{\delta_v}\,v_p^2 \, .
\label{turbulentdragbalance}
\end{equation}
Consequently,
\begin{equation}
v_p=\mu_p\,\left(\dfrac{2\,\beta\, (T_b-T_c)\,\mathcal{L}}{C_p} \, \dfrac{\rho_c\,\rho_{ice}\,g\,L\,\sin \theta\,
v_m}{\rho_b^2} \right)^{1/3} \, ,
\label{vpth}
\end{equation} 
 where $\mu_p=(f_D)^{-1/3}=0.2+0.38\, \cos^2 \theta$ 
from Chaigne, \textit{et al.}~\cite{Chaigne2023}. 
 Typically, for $T_b=20\,^\circ$C and $\theta=26.5^\circ$, $v_p$ is of the order of a few millimeters per second. 
 This order of magnitude enables us to neglect the contribution of the meltwater to the water flow when determining $v_p$. The incoming flow rate due to melting per width unit is indeed $v_m\, L$, whereas the flow rate of the time average current flowing through a thickness $\delta_v$ below the block is $v_p\, \delta_v$. Because $\delta_v$ is of order $L$, the ratio between these flows is $(v_m\, L)/ (v_p\, \delta_v)\approx v_m/v_p \approx 1/100$. Consequently, the contribution of the meltwater to the water flow can be neglected { when determining $v_p$}. 

Next, we perform momentum balance on the control volume shown in the schematics of Fig.~\ref{MeltingSchemaA}(a) to estimate the propulsion of the floating ice block~\cite{Chaigne2023},
\begin{equation}
    F_c \approx \rho_b \frac{W}{2}\,\delta_{x}\,v_p^2\,\sin \theta \,\cos\theta\, ,
    \label{Fc}
\end{equation}
where $\delta_x$ is the length  over which the flow is ejected. PIV measurements performed by Chaigne~\textit{et al.}~\cite{Chaigne2023}  with dissolving bodies have shown that the flow is mostly confined over a region with a thickness $\delta_v$ and width $W/2$. \blue{The thickness $\delta_v$ of this region, which exhibits significant velocity fluctuations, is of the same order as the block length, $L$, and far exceeds the viscous boundary layer for the velocity field.} Moreover, the same study has determined empirically a correlation between $\delta_x$ and the surface inclination $\theta$, $\delta_x=\frac{1}{2}\, \cos^2 \theta \, {L}$. We assume these flow characteristics to carry over here as well. Indeed, it can be noted from shadowgraph images shown in Fig.~\ref{fig:ice}(d) that $\delta_x$ is of order the block largest length $L$.\\ 

When the ice block reaches terminal velocity, the propulsion force is balanced by an inertial drag,
\begin{equation}
F_d=\dfrac{1}{2}\, C_d \, \rho_b \,L_A\,W\,U_b^2,
\label{Fdrag}
\end{equation}
where $C_d$ is the drag coefficient of the ice block, and ${L}_A \approx {L} \sin \theta $ is the projected length. 
Hence, we estimate the terminal velocity from the balance $F_c=F_d$, 
\begin{equation}
U_{b,th}=\sqrt{\dfrac{\cos\theta\, \delta_x}{\, C_d\, {L}}}\, v_p,
\label{Ubth}
\end{equation}
where $C_d$ is the drag coefficient of the ice block. We assume quadratic drag because the Reynods number $Re=\dfrac{L\,U_b}{\nu} \approx 600$ in our experiments and the flow below the block appears detached below the blocks in the shadowgraph images. These flow conditions are similar to those studied previously in the dissolving case~\cite{Chaigne2023}, where a quadratic drag was seen to be consistent with the observed kinematics. Following Ref.~\cite{Chaigne2023}, here we choose $C_d=0.6$. The variability in shapes and dimensions of the inclined ice blocks used in this study \blue{leads} to a significant uncertainty on the coefficient $C_d$, that we estimate to be of order $\pm 0.3$. Further, a skin friction can be present on the lateral side and needs to be taken into account as discussed in Appendix~\ref{Skinfriction}. In practice, we find this contribution is significantly smaller than the inertial drag over the range of ice block dimensions and velocity investigated. Accounting for skin friction results in an effective drag friction coefficient $C_d\approx 0.8$, within the range of $C_d \approx 0.6 \pm 0.3$. 
Substituting $v_p$ calculated using Eq.~(\ref{vpth}) in Eq.~(\ref{Ubth}), we can calculate the ice block velocity for the various experiments.

\subsection{First-order estimate of the terminal velocity of the ice block and physical remarks}
\label{scalings}
 To obtain a more general estimate of order one of the terminal velocity $U_{b,sc}$, we remove the angular dependency and estimate $v_m$ by approximating the Stefan condition given by Eq.~(\ref{Stefancond}) as 
\begin{equation}
v_m \sim \dfrac{\rho_i} {\rho_{ice}} \dfrac{C_p\, \kappa}{\mathcal{L}} \, \dfrac{(T_b-T_i)}{\delta_T}=  \dfrac{\rho_i} {\rho_{ice}} \dfrac{C_p\, \kappa}{\mathcal{L}} \, \dfrac{(T_b-T_c)}{\delta_i} \, .
\end{equation}
Substituting in Eq.~(\ref{vpth}) and noting that to first order $U_b \sim v_p$, we obtain,
 \begin{equation}
 \label{Eqscalinglaw}
U_{b,sc} \sim
  \beta^{1/3} (T_b-T_c)^{2/3}  
 \left(\dfrac{\rho_i \, \rho_c}{\rho_b^2} \right)^{1/3}
 \left( \dfrac{{L}}{\delta_i} \right)^{1/3} (\kappa \, g)^{1/3}.
\end{equation}
It can be noted that $\rho_{ice}$ is absent, implying that the ice block velocity does not depend on the ice density. Thus, the presence of trapped bubbles, {which changes the overall block density}, cannot be expected to significantly affect $U_b$. As we discuss later, even though the buoyancy of the released bubbles can affect the flow below the melting block, thus the melting and the propulsion~\cite{wengrove2023melting}, a notable difference between measured \blue{terminal velocities} $U_b$ of ice blocks with and without bubbles are not observed.
Further, for $T_b=20^\circ$C and ${L} \approx 10$~cm, $ \beta\, (T_b-T_c)^2 $ is of order $1.81 \times 10^{-3}$,  $\rho(T_i)\, \rho_c/\rho_b^2$ is approximately equal to $1$, and the ratio of length $L/\delta_i$ is roughly about $50$. The only dimensional factor is the characteristic velocity $(\kappa \, g)^{1/3} \approx 11$ mm s$^{-1}$. Substituting these estimates in Eq.~(\ref{Eqscalinglaw}), we obtain the terminal velocity $U_{b,sc} \,\blue{\approx} \,5$~mm\,s$^{-1}$, which is the same order of magnitude as in our experiments.  By combining Eq.~(\ref{deltaT}) and Eq.~(\ref{Eqscalinglaw}), we find, $U_{b,sc} \, \blue{\propto} \,  (T_b-T_c)^{8/9}$. The terminal ice block velocity is thus nearly proportional to $(T_b-T_c)$, and increases with the bath temperature $T_b$ as expected. 
Further, $U_{b,sc}$ increases with the block length as $L^{1/3}$ according to Eq.~(\ref{Eqscalinglaw}), implying a weak increase with block size. 

{An important remark pertains to the role of latent heat $\mathcal{L}$ on the propulsion, which is absent from Eq.~(\ref{Eqscalinglaw}). According to Eq.~(\ref{Ubth}), $U_b$ is proportional to $v_p$, which is computed using Eq.~(\ref{vpth}). In that equation, the latent heat $\mathcal{L}$ is multiplied by the melting velocity $v_m$, which according to Eq.~(\ref{Stefancond}), is proportional to  $1/\mathcal{L}$ in the limit of negligible heat flux in the ice, \textit{i.e.} when the heat flux to raise the block temperature is relatively small compared to the latent heat ($St \ll 1$). Therefore, the magnitude of the latent heat $\mathcal{L}$ in our experiments does not influence the value of the propulsion velocity. By contrast, the lifetime of the melting block is directly related to the magnitude of latent heat, because $v_m$ is inversely proportional to $\mathcal{L}$.}

Moreover, in modeling our experiments, we neglected the effect of propulsion flow on melting dynamics, as the propulsion velocities are too small to effectively shear the thermal boundary layer and enhance the melting rate. The Richardson number $Ri$ compares the magnitude of the buoyancy force to the inertial force caused by the flow,
\begin{equation}
Ri=\dfrac{\Delta \rho \, g\, \mathcal{D}}{\rho \mathcal{V}^2}\, ,
    \label{Richardsonnumber}
\end{equation}
where $ \mathcal{D}$ and $\mathcal{V}$ are characteristic length and velocity, respectively. In our experiments, $ \mathcal{D}=L$, $ \mathcal{V} = v_p \approx U_b$, and  $\Delta \rho /\rho= \beta \, (T_b-T_c)^2$ according to Eq.~(\ref{rhoeau}). When $T_b = 20\,^\circ$C, $L=0.2$\,m and $U_b=5$~mm s$^{-1}$, $Ri \approx 140$. This large $Ri$ means that the buoyancy force is dominant compared to the shear force  
due to the block motion and the gravity driven current $\bm{v}$, justifying that the effect of the ice block velocity on the melting velocity is negligible.

\subsection{Comparison with measured ice block velocity}
\label{ParametricStudy}

\begin{figure}[t]
    \centering
    \includegraphics[height=9cm]{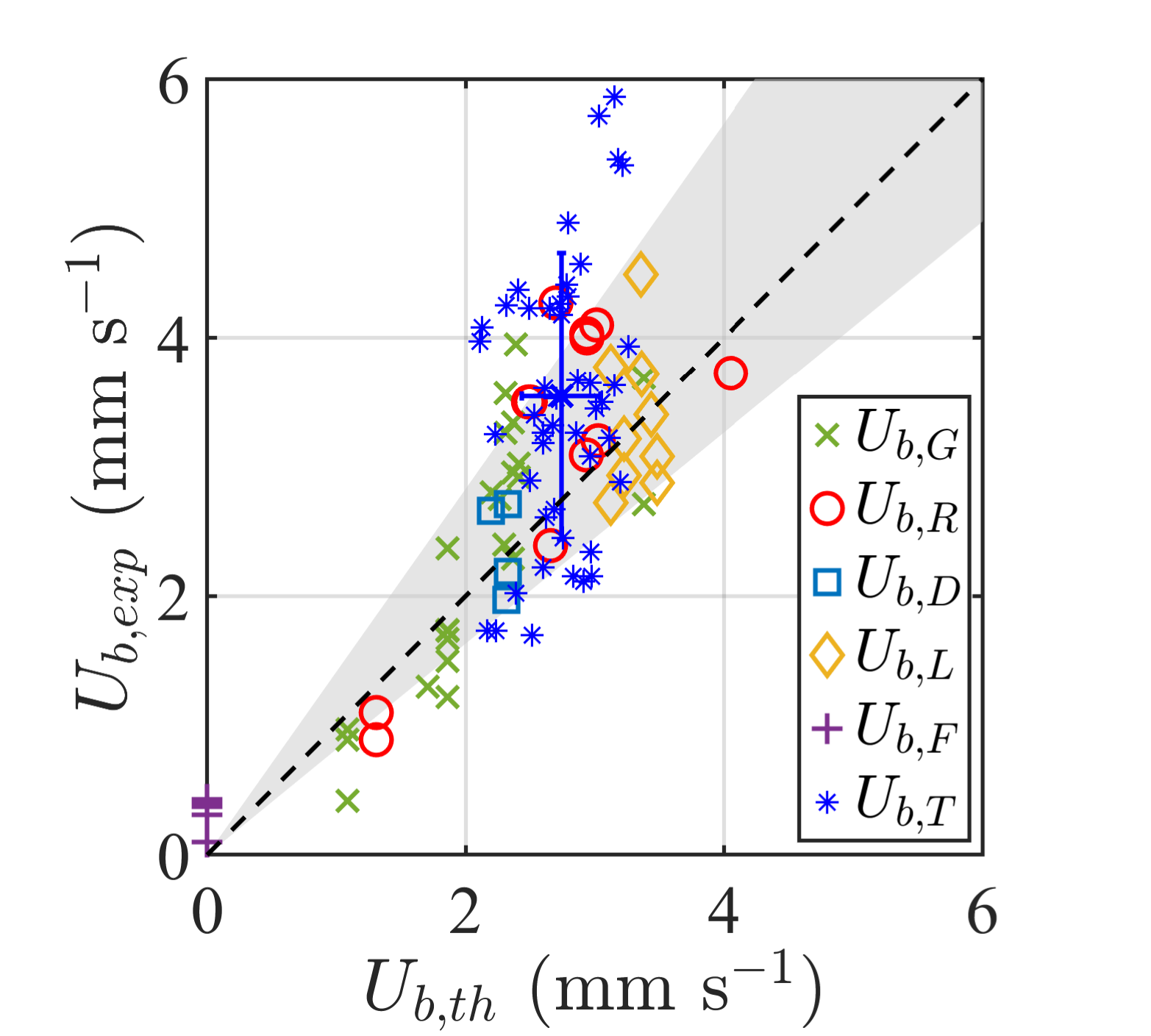}
    \caption{ \blue{Comparison of the measured terminal velocities  $U_{b,exp}$ with the theoretical estimates $U_{b,th}$ in a fresh water bath. The data} fall along a line with slope one (dashed line), showing that they are consistent within experimental and model parameter dispersion. The experimental parameters can be found in Appendix~\ref{Datasets}.     $U_{b,T}$ correspond to measurements with top view imaging. The error bar \blue{corresponds to one standard deviation around the mean value and indicates the dispersion for this data set}. The other data sets correspond to shadowgraph experiments with side view imaging where the lateral motion of the block is constrained by wires. Specifically, $U_{b,D}$, $U_{b,L}$ and $U_{b,F}$ are made of nearly clear ice and have a low concentration of trapped air bubbles. $U_{b,F}$ are symmetric rectangular ice blocks, for which the measured propulsion velocity is close to zero. The gray area denotes the variation of $U_{b,th}$ while varying $C_d = 0.6 \pm 0.3$.} 
      \label{fig:UbexpvsUbtheo}
\end{figure}

Fig.~\ref{fig:UbexpvsUbtheo} shows a scatter plot of the measured ice block velocity $U_{b,exp}$ versus the theoretical value $U_{b,th}$ according to Eq.~(\ref{Ubth}) with the various block shapes and bath temperature in our study (see Appendix~\ref{Datasets}). The measurements include those in which the ice block is constrained to move along the $x$-axis between two wires and viewed from the side with shadowgraph imaging, and those in which the ice block is not constrained laterally and obtained with top view imaging to find the magnitude of the velocity from the horizontal components. Fig.~\ref{fig:UbexpvsUbtheo} shows that the points are located close to the dashed line assuming $C_d=0.6$, and fall mostly within a narrow gray band bounded by $C_d = 0.3$ at the top and $C_d = 0.9$ at the bottom.  
The points are less scattered in the case where motion is constrained to one-dimension compared those obtained while the block is unconstrained. This leads to greater directional stability because rotation is inhibited, and hence to closer agreement. 
  
To explore the influence of the main experimental parameters and compare the measurements and predictions while several experimental parameters are varied, we rescale the block speeds to include the expected theoretical dependency of parameters which are not being examined.  As reference parameters, we choose length $L_r=0.1$\,m, inclination $\theta_r=\mathrm{atan}(5/10)\approx 26.6^\circ$, and bath temperature $T_{b,r}=20\,^\circ$C.

\begin{figure}[t!]
    \centering
    \includegraphics[height=8cm]{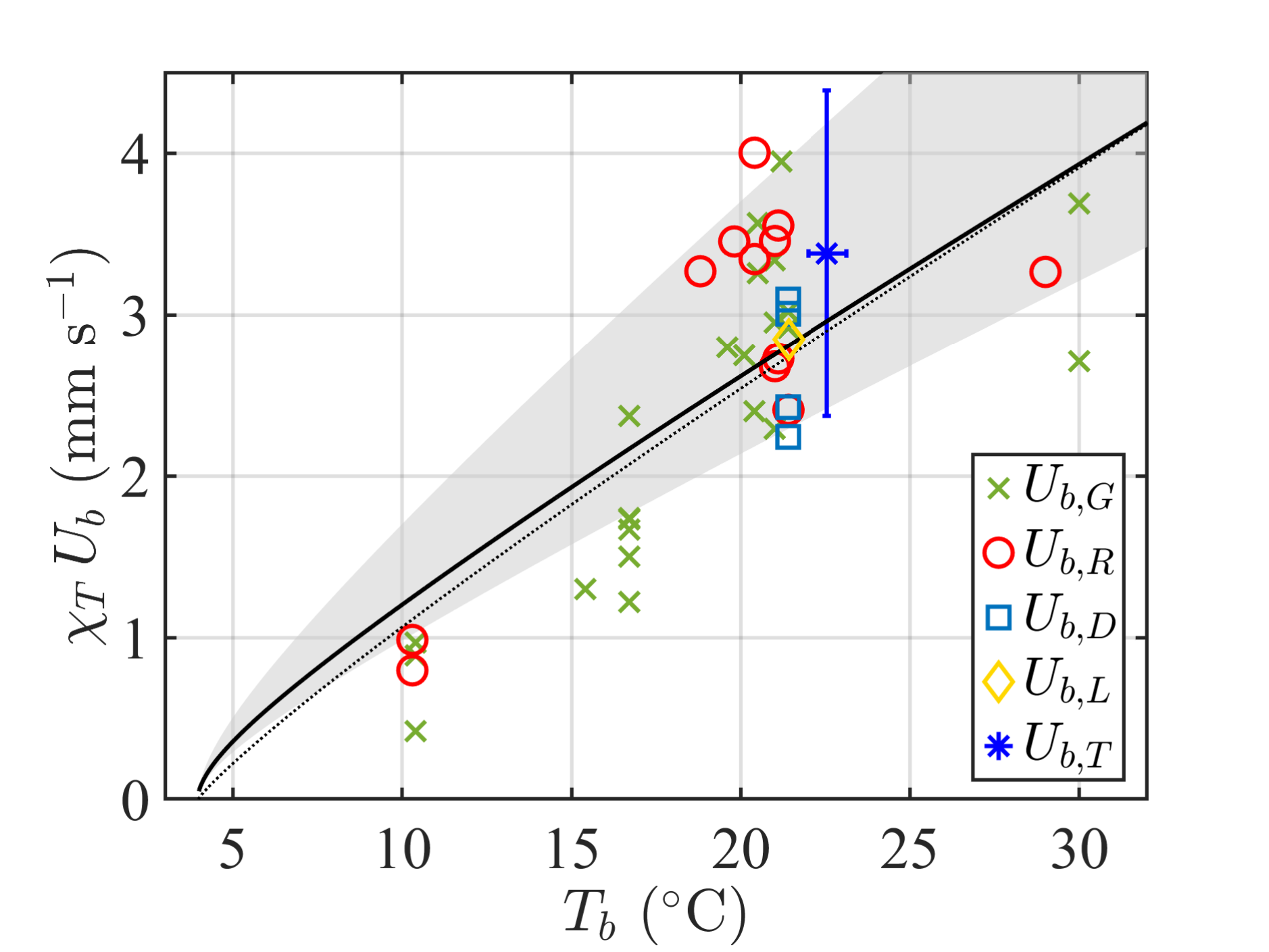} 
               \caption{Rescaled measured terminal velocity $\chi_T \, U_{b,exp}$ as a function of the bath temperature $T_b$ and comparison with model $U_{b,th}$ (black line). The measured velocities are rescaled by a factor $\chi_T$ to compare experiments with different system properties (see text). 
               \blue{The experimental parameters can be found in Appendix~\ref{Datasets}. $U_{b,T}$ correspond to measurements with top view imaging. The error bar corresponds to one standard deviation around the mean value and indicates the dispersion for the measurements view from the top. The other data sets correspond to shadowgraph experiments with side view imaging where the lateral motion of the block is constrained by wires. Specifically, $U_{b,D}$ and $U_{b,L}$ are made of nearly clear ice.}
               The gray area denotes the variation of $U_{b,th}$ while varying $C_d = 0.6 \pm 0.3$. The dot line corresponds to the scaling law $U_{b,sc} \,\blue{\propto}\, (T_b-T_c)^{8/9}$.}
    \label{fig:UbexpvsT}
\end{figure}

\begin{figure}[t!]
    \centering
        \includegraphics[height=8cm]{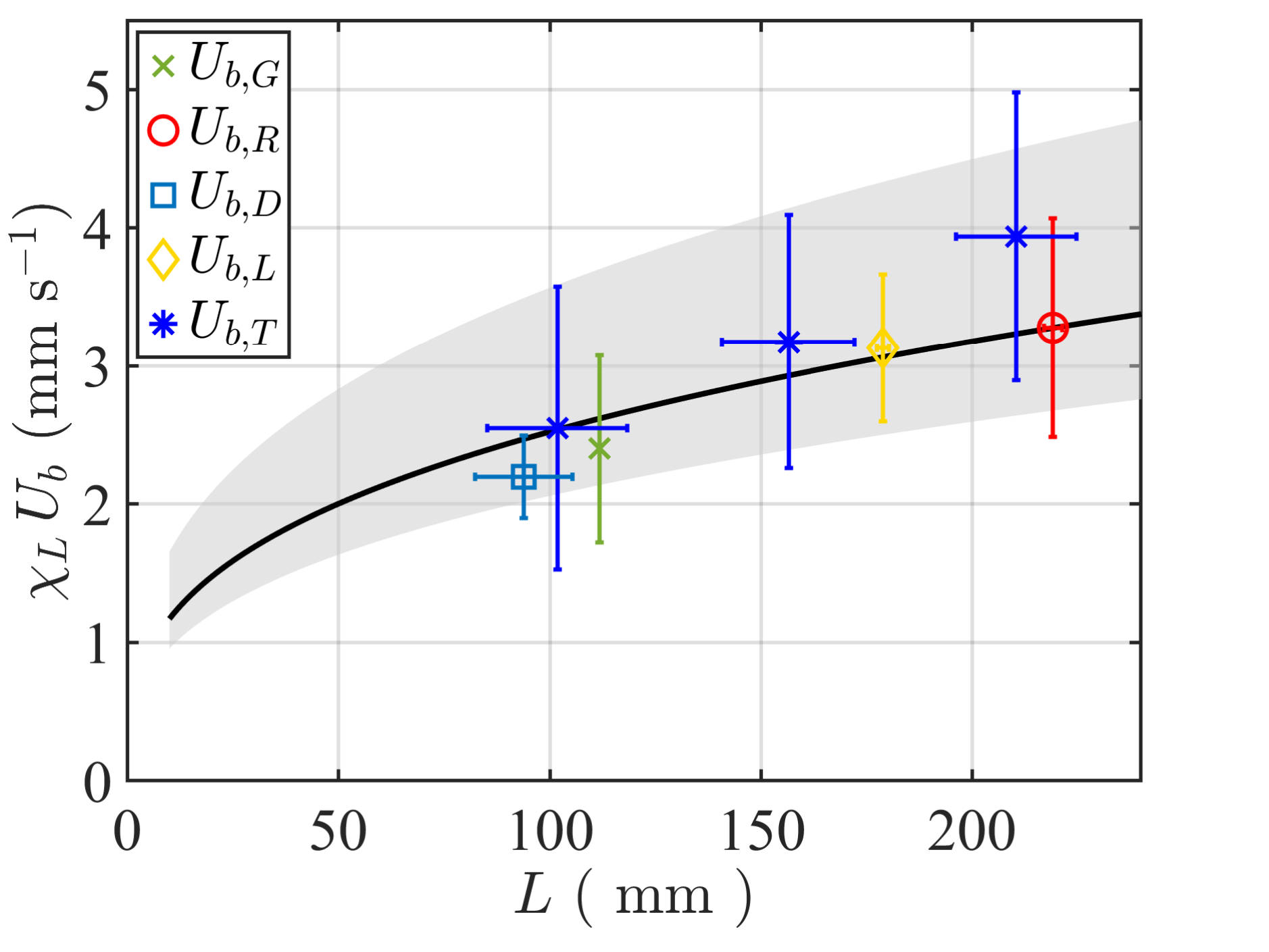}         \caption{Rescaled measured terminal velocity $\chi_L \, U_{b,exp}$ as a function of the  ice block length $L$ and comparison with model $U_{b,th}$ (black line). The measured velocities are rescaled by a factor $\chi_L$ to compare experiments with different system properties (see text). \blue{The experimental parameters can be found in Appendix~\ref{Datasets}. Specifically, $U_{b,D}$ and $U_{b,L}$ are made of nearly clear ice. The dataset for $U_{b,T}$ (view from the top) is divided into three subranges in $L$ to better visualize the influence of $L$, while maintaining statistical averaging. The error bars show the average value and $\pm$ the standard deviation indicating the dispersion of measurements inside a dataset}. The gray area denotes the variation of $U_{b,th}$ while varying $C_d = 0.6 \pm 0.3$. By construction, $U_{b,th}$ follows exactly the scaling $U_{b,th} \,\blue{\propto}\, L^{1/3}$.}
    \label{fig:UbexpvsL}
\end{figure}

\begin{figure}[h!]
    \centering
 \includegraphics[height=8cm]{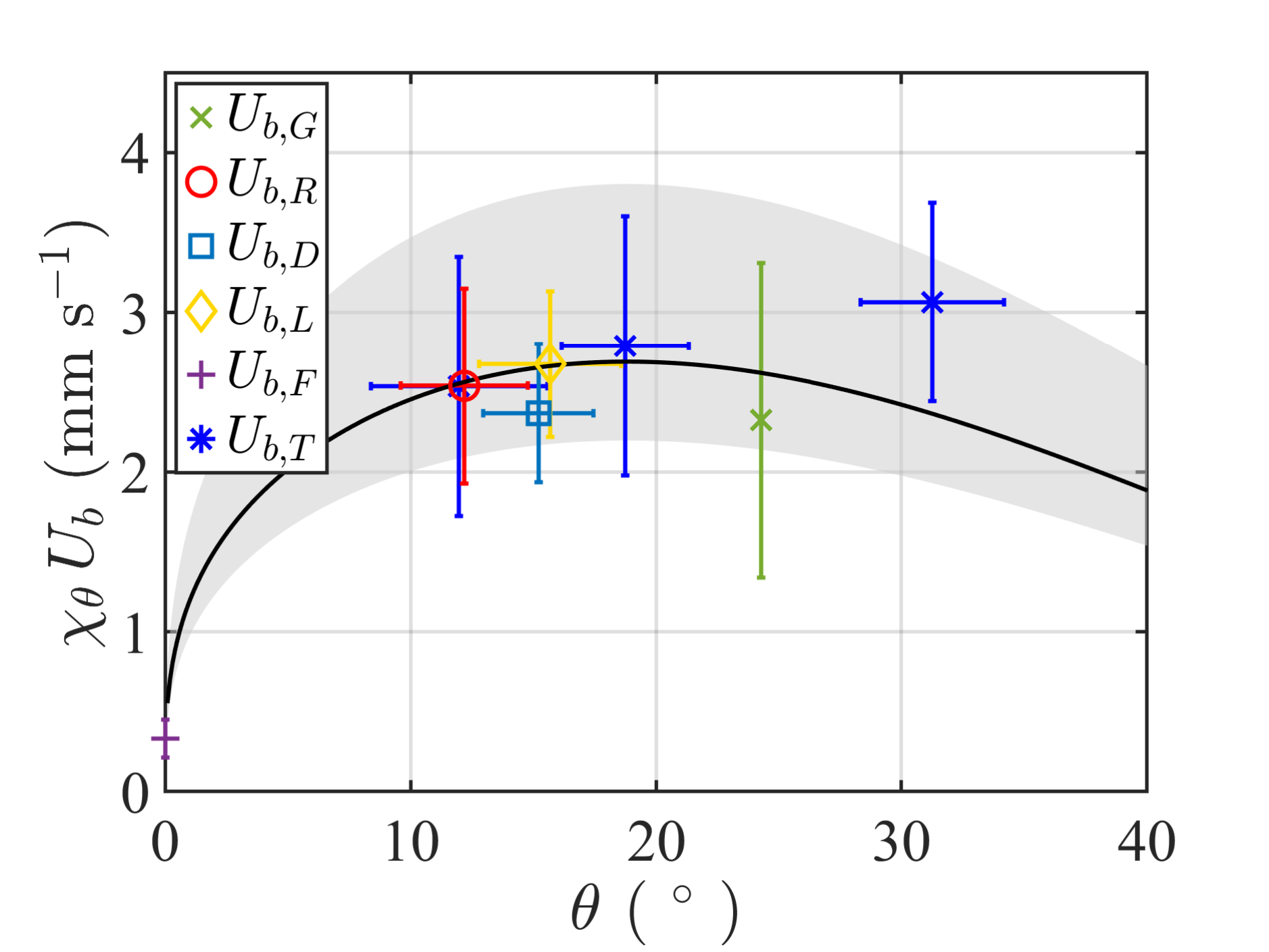}
\caption{Rescaled measured terminal velocity $\chi_\theta\,U_{b,exp}$ as a function of the ice block inclination $\theta$ and comparison with model $U_{b,th}$ (black line). The measured velocities are rescaled by a factor $\chi_\theta$ to compare experiments with different system properties (see text). \blue{The experimental parameters can be found in Appendix~\ref{Datasets}. Specifically, $U_{b,D}$ and $U_{b,L}$ are made of nearly clear ice. The dataset for $U_{b,T}$ (view from the top) is divided into three sub-ranges in $\theta$ to better visualize the influence of $\theta$, while maintaining statistical averaging. The other datasets have been also averaged with $\theta$, due to their restricted range of inclination values in a given dataset. The error bars show the average value and $\pm$ the standard deviation indicating the measurement dispersion. The gray area denotes the variation of $U_{b,th}$ while varying $C_d = 0.6 \pm 0.3$. We note for the data set $\textbf{F}$ ($U_{b,F}$) for which $\theta\approx 0^\circ$, that $U_b\approx 0$.} For the other data sets, we observe a weak dependency with $\theta$ in agreement with the theoretical model.}
    \label{UbexpvsThet}
\end{figure}

The effect of \blue{the bath temperature} $T_b$ is examined by multiplying the measured \blue{terminal velocities} $U_b$ by a factor $\chi_T  = \dfrac{\Theta(\theta_r)}{\Theta(\theta)} \left(\dfrac{L_r}{L}\right)^{1/3} $, where
\begin{eqnarray}
\Theta(\theta)&=& \left( \dfrac{ \cos^2\theta \, \sin(2 \theta)}{\sin\theta}   \right)^{1/2} \,   (0.2+0.38\,\cos^2 \theta )  (\sin \theta)^{1/3} \, (\cos\theta)^{1/9} \nonumber, \,\, \blue{\rm or} \\
\blue{\Theta(\theta)}&\blue{=}&\blue{    (0.28+0.54\,\cos^2 \theta )  (\sin \theta)^{1/3} \, (\cos\theta)^{29/18}} \, ,
\label{Inclinationdep}
\end{eqnarray}
is obtained by combining Eqs.~(\ref{vmelting}), (\ref{vpth}), and (\ref{Ubth}). It can be noted that this angular dependence is by construction identical to the one found for the dissolving boats~\cite{Chaigne2023}. The factor $(L_r/L)^{1/3}$ arises from the scaling law between $U_b$ and $L$ in Eq.~(\ref{Eqscalinglaw}). Fig.~\ref{fig:UbexpvsT} shows the rescaled $U_{b}$ 
plotted as a function of \blue{the bath temperature} $T_b$ over the measured range 10$\,^\circ$C and 30$\,^\circ$C. The increase of $U_{b}$ with $T_b$ is consistent with the theoretical estimate $U_{b,th}$ obtained from Eq.~(\ref{Ubth}) with the references length $L_r$ and inclination $\theta_r$. Further, one can see that it follows the approximate scaling $U_{b,sc} \,\blue{\propto}\, (T_b-T_c)^{8/9}$ developed in Sec.~\ref{scalings}.

The comparison of the rescaled values of $U_b$ as a function of the inclined length $L$ with those calculated using Eq.~(\ref{Ubth}) is shown in Fig.~\ref{fig:UbexpvsL}. Here, $U_{b,exp}$ has been multiplied by $\chi_L =\dfrac{\Theta(\theta_r)}{\Theta(\theta)} \, \left(\dfrac{T_{b,r}-T_c}{T_b-T_c}\right)^{8/9} $, whereas $U_{b,th}$ is plotted for the fixed reference values of the parameters $T_{b,r}$ and $\theta_r$ and variable $L$. The experimental data are consistent with the slow increase with $L$, which is predicted theoretically.

Finally, we compare the rescaled values of $U_{b,exp}$ as a function of the inclination of the bottom surface of the ice block $\theta$ in Fig.~\ref{UbexpvsThet} to the model $U_{b,th}$. Here, $U_{b,exp}$ is multiplied by $\chi_\theta = \left(\dfrac{L_r}{L}\right)^{1/3} \left(\dfrac{T_{b,r}-T_c}{T_b-T_c}\right)^{8/9}$, whereas $U_{b,th}$ is plotted for the fixed reference values of the parameters $T_{b,r}$ and $L_r$ and variable $\theta$. In the range of inclinations between $12$ and $32^\circ$, the data are also consistent with the theoretical predictions. We observe also a nearly zero velocity for the symmetric blocks. 

Thus, we find that the estimated $U_b$ trends according to our model in terms of $T_b$, $L$ and $\theta$ are in overall agreement with the data considering the measurement variations, and the number of approximations that were needed in deriving Eq.~(\ref{Ubth}). 

\section{Melting driven-propulsion in salt water} 
\label{saltwater}

We now investigate the ice blocks move in saltwater baths and their direction by varying its salinity $s_A$ from 0 to $35$\,g\,kg$^{-1}$, the salinity of ocean water. 
The physics of ice melting in oceans is considerably different from that in freshwater~\cite{McCutchan2022,du2024physics}. Besides decreasing the melting temperature $T_m$, salinity changes the water density more strongly in comparison with temperature. Upward plumes of fresh water have been reported in the vicinity of icebergs~\cite{cenedese2023,nash2024turbulent} and ice shelves~\cite{jenkins2010observation,davis2019turbulence,Hewitt2020} as a result of it being less dense than the surrounding saltwater.
Fig.~\ref{fig:salinity}(a) shows a plot \blue{of the rescaled measured terminal velocity $\chi_s\,U_b$ as a function of salinity $s_A$, where $\chi_s$ is a dimensionless rescaling factor. We report a systematic decrease of $\chi_s\,U_b$ over the entire range of $s_A$.} The direction of motion remains the same as in freshwater as $s_A$ is increased from 0 to that of ocean water. This is counterintuitive considering the fact that the motion of the ice block is opposite to the flow in freshwater, and meltwater which has zero salinity rises in saltwater. This alerts us to the fact that the explanation for ice block propulsion is more subtle in saltwater. \blue{In order to compare runs operated with variable parameters, we have introduced the rescaling factor $\chi_s=\left(\dfrac{L_r}{L}\right)^{1/3}\,\dfrac{\Theta(\theta_r)}{\Theta(\theta)} \, \left(\dfrac{T_{b,r}-T_c}{T_b-T_c}\right)^{8/9}$, where $L_r=210$ mm, $\theta_r=25^\circ$ and $T_{b,r}=23\,^\circ$C corresponding approximately to the measurements performed in saltwater of data set \textbf{T}. The use of the freshwater rescaling laws obtained in section~\ref{scalings} in saltwater are not fully justified. However, as it will be shown later, in the range of tested parameters, the mechanism described in section \ref{sec:PropulsionModel} applied also in salt water in first approximation. Moreover, the variation range of parameters for the experiments shown in Fig.~\ref{fig:salinity}(a) remains moderate: $17.8<\theta<26.6^\circ$, $112<L<267$ mm and $19.6<T_b< 28.5 \,^\circ$C.} 

Fig.~\ref{fig:salinity}(b) shows a shadowgraph image towards probing the flows below the floating ice block (see also Movie~S4~\cite{sup-doc}). The upward flow of the meltwater adjacent to the ice block because of its buoyancy is not discernible from these images. Rather, descending plumes are observed, similar to those in freshwater shown in Fig.~\ref{fig:ice}(d), showing that the cooling of the bath water below the ice block causes it to sink. 
To find the meltwater and its flow, we performed complementary experiments with ice blocks that were dyed with red food coloring. The resulting images in freshwater and saltwater are shown in Fig.~\ref{fig:dye}. The sinking plumes are colored by the dye in the freshwater bath, consistent with shadowgraph images in freshwater. In contrast, the dye is observed to spread at the free surface on top of the saltwater bath, and the liquid below the block remains colorless. Thus, observations using dye and shadowgraph imaging confirm that freshwater released by melting flows upward while remaining in close contact with the block, whereas cooling of the bath generates a detached convection flow, as revealed by the shadowgraph images. Hence the illustrations appear to show that the contribution of convection due to the cooling of the surrounding salt bath dominates the rising cold fresh water near the melting surface, giving rise to propulsion in the same direction as in freshwater.

\begin{figure}[t]
    \centering
    \includegraphics[width=0.8\linewidth]{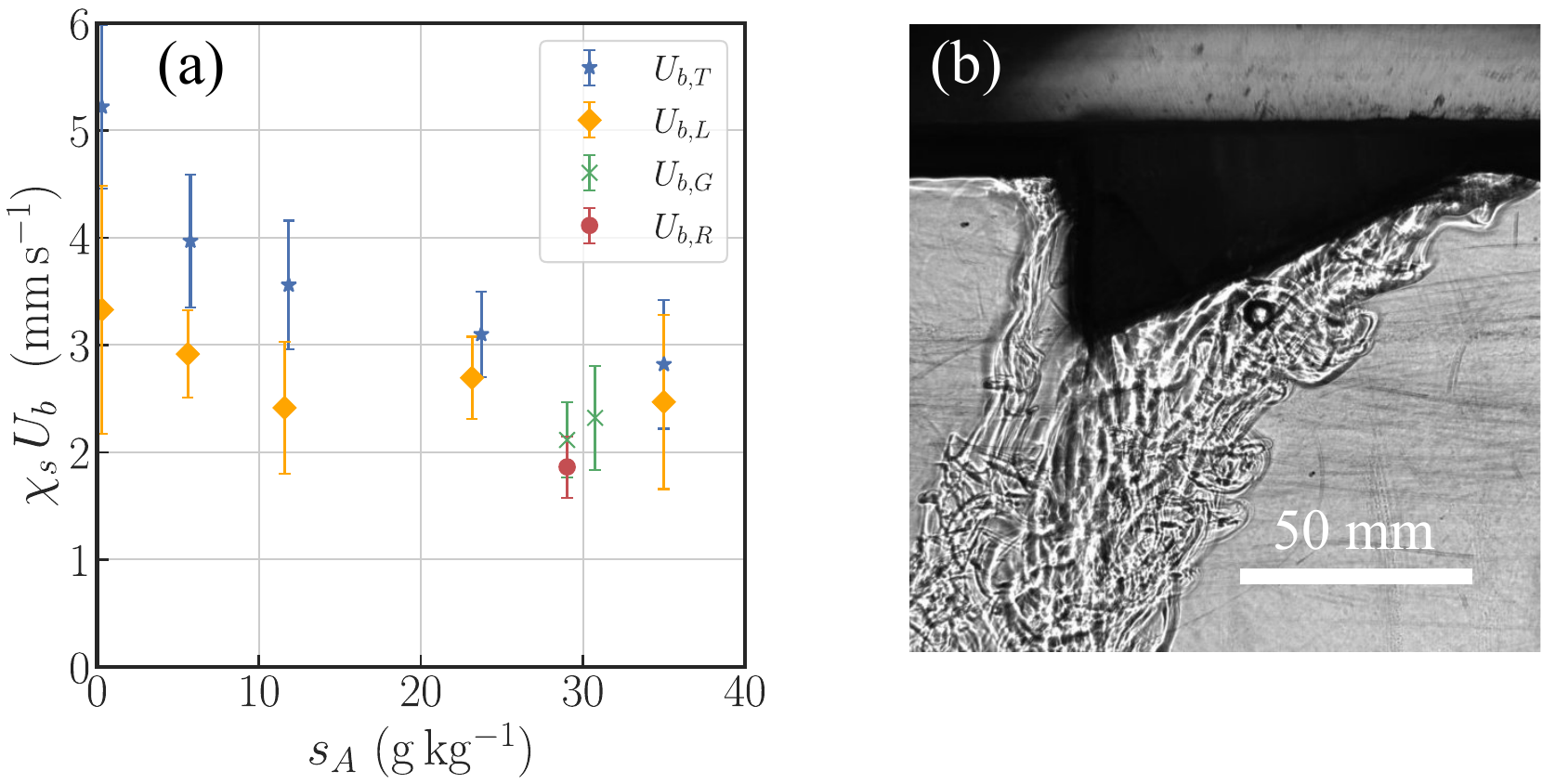}
    \caption{ \blue{Melting-driven propulsion in a saltwater bath. (a) The rescaled measured terminal velocity $\chi_s\,U_{b}$ as a function of the bath salinity $s_A$. The experimental parameters for the different datasets can be found in Appendix A. $U_{b,T}$ corresponds to measurements with top view imaging and the others were obtained from shadowgraph measurements. $U_{b,L}$ is obtained with nearly clear ice blocks. 
    The measurements obtained with various parameters have been rescaled by a factor $\chi_s$ referring to the measurements $U_{b,T}$, $L = 210$\,mm, $\theta = 25^\circ$, $T_b = 23\,^\circ$C (see text). The errorbars  correspond to one standard deviation around the mean value and indicate the dispersion of measurements for each data set and salinity value}. (b) Shadowgraph image of an ice block moving in salt water ($s_A = 31$ g kg$^{-1}$, $L_h \times W\times H\approx 100\times 100 \times 50$ mm$^3$, $T_b=24.6\,^\circ$C, $U_{b}=2.28$\,mm\,s$^{-1}$).  (See Movie S4 in SM). }
    \label{fig:salinity}
\end{figure}

\begin{figure}
    \centering
    \includegraphics[width=10cm]{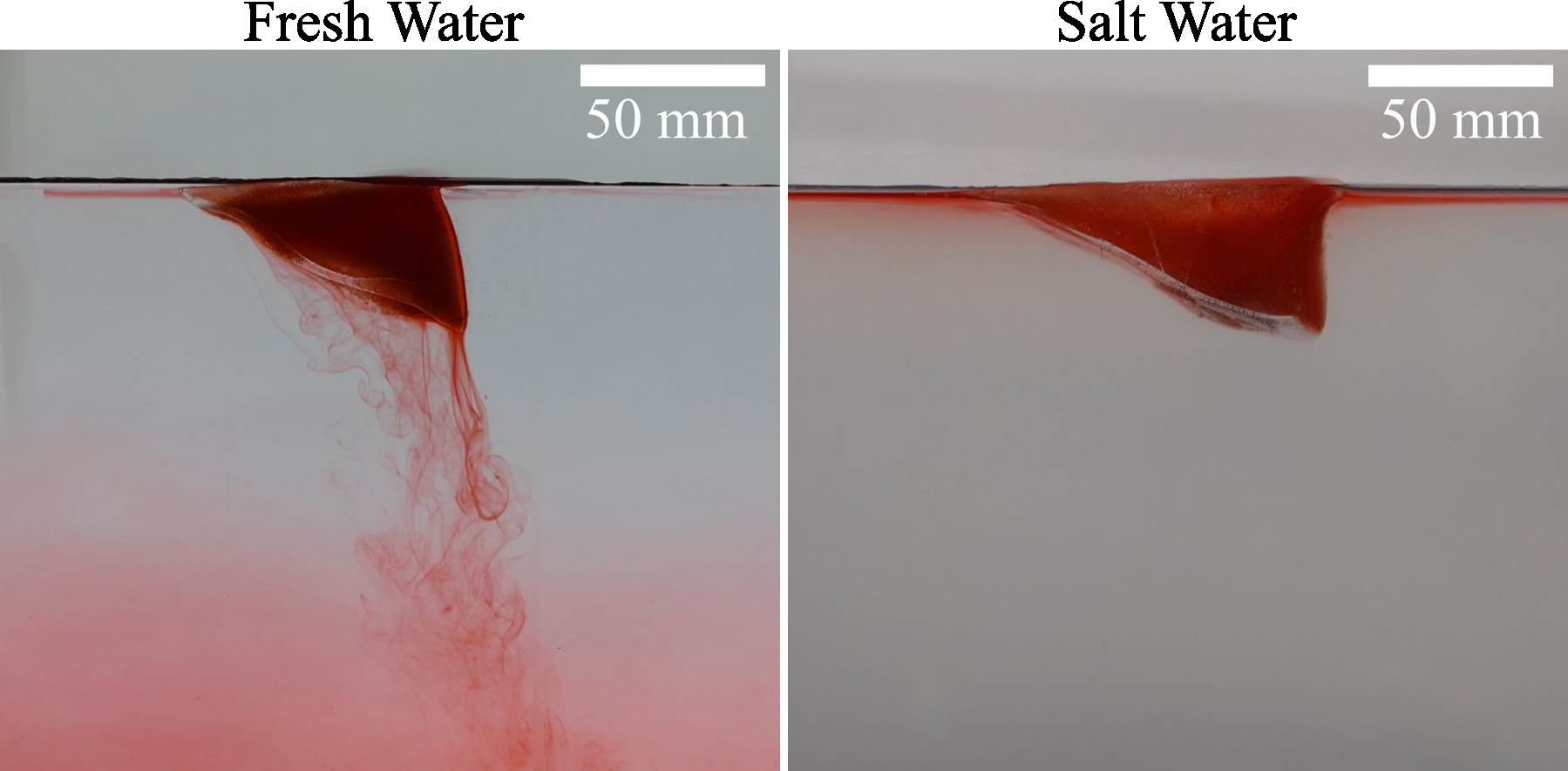}
    \caption{\blue{Evidence of meltwater flow.} An ice block ($L_h \times W\times H\approx 121\times 80 \times 50$ mm$^3$ and $\theta\approx 25^\circ$)  dyed with red food color to track melt water in clear fresh water and in salt water ($35$\,g salt per kg of water). The snapshots were taken about 270 seconds after placing the ice block in the bath ($T_b\approx 22\,^\circ$C). The melt water descends to the bottom in fresh water and rises to the surface in salt water because of the relative density difference with the bath.}
    \label{fig:dye}
\end{figure}

\begin{figure}[h]
    \centering
    \includegraphics[width=0.6\textwidth]{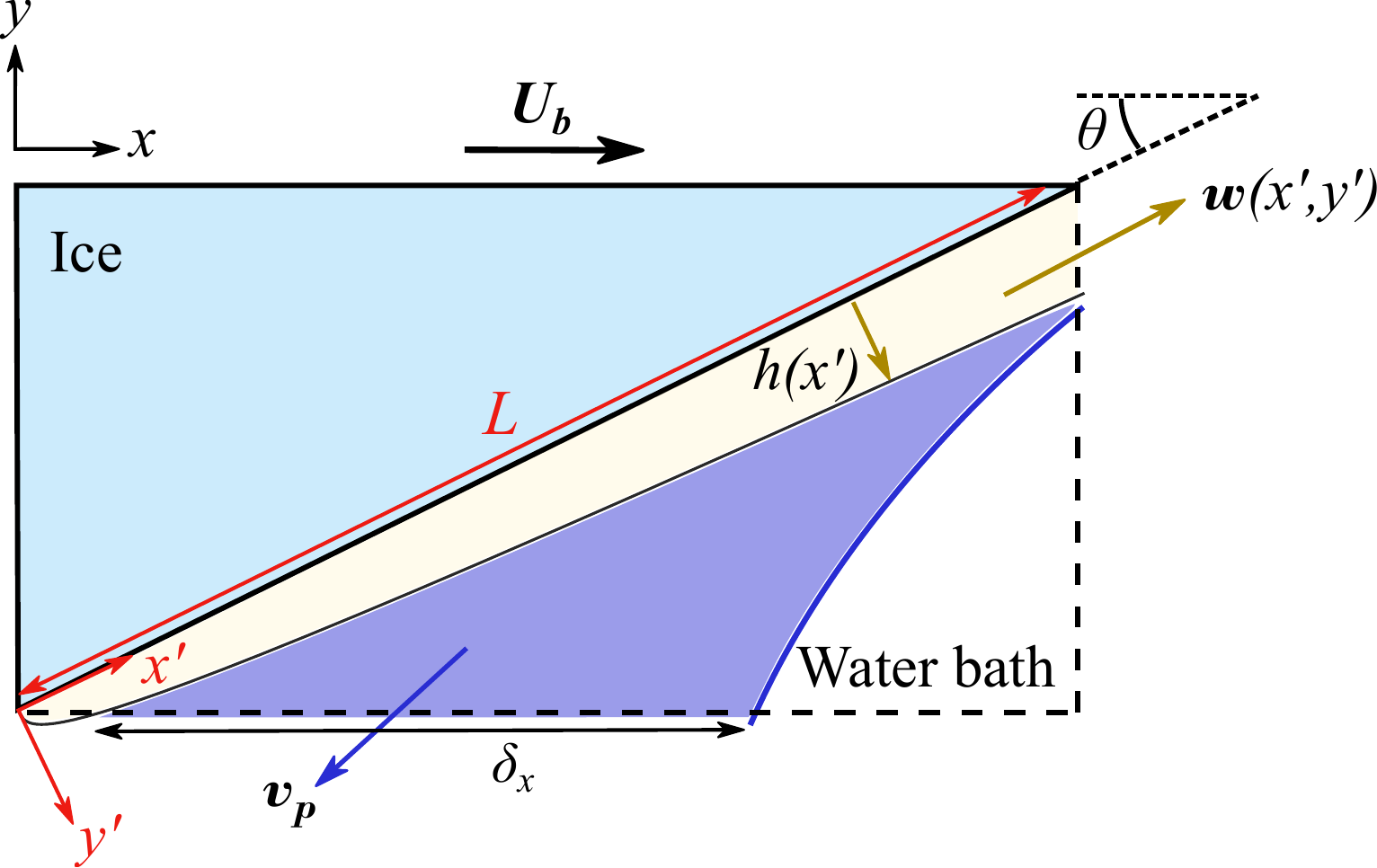}
    \caption{A schematic of a melting ice block moving while floating in salt water with velocity $\bm{U_b}$.  Two layers of fluid coexist below the ice block and flow in the opposite directions. 
    A thin layer of pure meltwater adjacent to the inclined surface of the ice block is accelerated upwards due to its lower density compared with saltwater. Its thickness $h$ and velocity $w$ increase along the $x'$ axis while aligned with the inclined surface.  A saltwater layer further away from the ice is cooled because of the heat drawn by the melting of the ice block. This denser fluid flows downwards because of gravity and escapes the control volume indicated by the dashed line over a length $\delta_x$ and a velocity $\bm{v_p}$.} 
    \label{SchemMeltLayer}
\end{figure}

\subsection{Propulsion model for ice melting in saltwater}
\label{ModelSalt}

As discussed in the introduction and recent reviews, the melting of fresh water ice bodies including icebergs, ice shelves, sea ice floes in the ocean is a complex subject~\cite{Malyarenko2020,McCutchan2022,du2024physics}. Several regimes have been identified to distinguish and describe ice melting in the ocean~\cite{Malyarenko2020}, depending on the ice surface orientation, the presence or absence of turbulence, the stratification of the salt concentration and the water temperature. The temperature of maximal density $T_c$ decreases with the salinity and is absent for a salinity above approximately $24$\,g~kg$^{-1}$, suppressing reverse buoyancy at low temperatures~\cite{du2024physics}.
Experimental studies have investigated the melting of ice block in a homogeneous saltwater bath in the absence of salt stratification at temperatures above 10$\,^\circ$C~\citep{russell1980melting,fukusako1994melting,yamada1997melting,fitzmaurice2017nonlinear,hester2021aspect,xu2024buoyancy,bellincioni2024melting}. According to Yamada, {\it et al.}~\cite{yamada1997melting}, a double diffusive convection flow occurs near immersed ice blocks under those conditions, which is driven by the sinking cooled salt water below the block and rising fresh meltwater above. We note also a non-monotonic dependence of the melting rate with the salinity has been reported in experiments, with different behavior at the side and the base of the ice block~\citep{russell1980melting}.

Based on these observations, we develop a simplified model of the propulsion of the asymmetric ice blocks in saltwater combining the fact that the meltwater rises and the bath water below sinks because of cooling. Except for the calculation of the melting velocity, the propulsion model is similar to 
the one presented for freshwater in Sec.~\ref{ModelMelting}  after incorporating  salinity into the density dependence and the physical parameters. The density of saltwater as a function of temperature and salinity, and the values of the physical parameters are obtained using the Gibbs-SeaWater (GSW) Oceanographic Toolbox of the Thermodynamic Equation of Seawater - 2010 (TEOS-10)~\citep{mcdougall2011getting}.

A schematic of the flows generated by melting and the cooling of the bath water based on observations is shown in Fig.~\ref{SchemMeltLayer}. As in the freshwater case, we consider the inverted inclined ice surface with $x'$ and $y'$ as the coordinates along and perpendicular to the ice block with the origin located at the bottom of the ice block. A layer of pure meltwater rises up the block while ice melting cools the saltwater layer below.
The cooling leads the saltwater in that layer to descent because of its increased density relative to the bathwater further below. Therefore, we assume that the melting velocity of the ice $v_m$ is determined by thermal convection. Because the ice is in contact with fresh meltwater, the ice melts at a the melting temperature for fresh water, \textit{i.e.} $T_i=T_m=0\,^\circ$C. As justified later in Section~\ref{Meltlayer}, we assume that the meltwater layer thickness is small compared to the thermal boundary layer $\delta_T$, and its temperature remains close to $T_m=0\,^\circ$C.}
As in Sec.~\ref{ModelMelting}, we calculate the thickness of the unstable boundary layer $\delta_i$, and derive, 
\begin{equation}
\delta_T=\dfrac{T_b-T_i}{T_b-T_{max}(s_A)}\,\delta_i\, , \quad \mathrm{with} \quad \delta_i=\left( \dfrac{Ra_c\, \kappa\,\nu}{g\,\cos \theta}\right)^{1/3} \, \left( \dfrac{\Delta \rho(s_A)}{\rho} \right)^{-1/3} \, ,
\label{deltaT2}
\end{equation}

where $T_{max}(s_A)$ is the temperature at which density of salt water at salinity $s_A$ is at its maximum over the range $[T_i,T_b]$, and $\Delta \rho = \rho(T_{max}(s_A),s_A)-\rho(T_b,s_A)$ is the density difference between salt water at $T_{max}$ and salt water at $T_b$. For $s_A \leq 18$\,g kg$^{-1}$, $T_{max}(s_A)=T_c(s_A)$ and $\delta_T$ is evaluated as in freshwater with Eq.~(\ref{deltaT}). When $s_A > 18$\,g kg$^{-1}$, $\rho(T,s_A)$ decreases monotonically between $T_i$ and $T_b$. Then, we assume $T_{max}(s_A)=T_i$, which gives $\delta_T=\delta_i$.

Using the Stefan condition Eq.~(\ref{Stefancond}), we obtain the melting velocity, 
\begin{equation}
v_m  = \Gamma \dfrac{\rho_i}{\rho_{ice}}\,\dfrac{c_p\,\kappa}{\mathcal{L}}\,\dfrac{T_b-T_i}{\delta_T} \, .
\label{Stefan2}
\end{equation}

A plot of $v_m$ versus $s_A$ for various $T_b$ between $1\,^\circ$C and $30\,^\circ$C is shown in Fig.~\ref{MeltingrateSalt}(a). We find that $v_m$ increases with $s_A$, because the density contrast between warm and cold water is enhanced by the salinity.} 
 
\begin{figure}[h]
    \centering
    \includegraphics[width=0.9\textwidth]{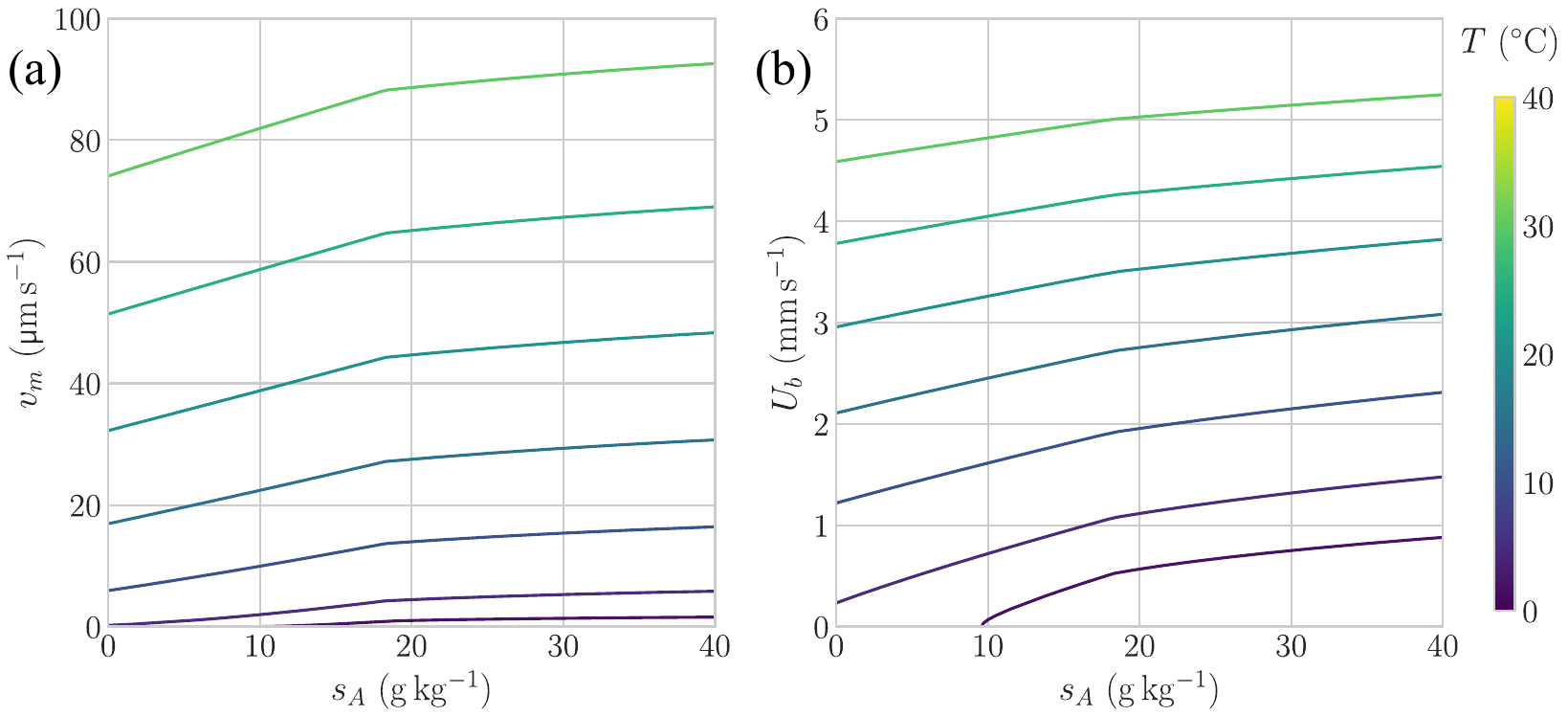}
    \caption{ \blue{Model predicting the melting velocity in salt water.} (a) The melting velocity evaluated using Eq.~(\ref{Stefan2}) increases with bath salinity for  $T_b$ between 0 and $30\,^{\circ}$C. The color bar maps the line color to $T_b$. 
    (b) The theoretical ice block velocity as a function of bath salinity at selected bath temperatures $T_b$ ($L=110$\,mm and $\theta = 23^\circ$).} 
    \label{MeltingrateSalt}
\end{figure}

{We obtain $v_p$ by substituting in Eq.~(\ref{Stefan2}) in Eq.~(\ref{vpth}). As in freshwater, $U_b$ is then obtained from Eq.~(\ref{Ubth}) by balancing 
the propulsion force with the inertial drag.} 
{Fig.~\ref{MeltingrateSalt}(b) shows that $U_b$ increases with salinity and bath temperature. For example, going from $s_A=0$ to $s_A=40$ g kg$^{-1}$ increases the predicted velocity by 25\% at $20\,^\circ$C. This is in clear contradiction with the experimental data, which suggest a significant decrease in $U_b$ with $s_A$. In fact one needs to take into account that as pure meltwater rises up along the block, it carries momentum in the opposite direction to the layer of cooled water.  
We calculate this contribution to the global momentum balance in the following. }  

\subsection{Melt layer dynamics}
\label{Meltlayer}

{We estimate the momentum carried by the melt layer by assuming a laminar convection flow driven by the density difference between meltwater with negligible salinity and the saltwater of the bath. The inclination of the ice block plays a crucial role allowing the meltwater flow to rise and avoiding the development of a stable insulating layer composed of fresh water between the bath and the ice. As shown in Fig.~\ref{SchemMeltLayer}, $u$ and $w$ denote the velocity components perpendicular and parallel to the block, respectively, and $h(x')$ the thickness of the melt layer.}  
Further, the edge of the melt layer $y'=h$ is assumed stress-free which leads the meltwater flow to have a half-Poiseuille (or Nusselt) profile,
\begin{equation}
w(x',y')=K\, y' \,(2\,h(x')-y'), \quad \mathrm{with} \quad K=\dfrac{g \, \sin \theta \Delta \rho}{2\, \rho\, \nu} \, ,
\label{eq:meltlayervelocity}
\end{equation} 
where $\Delta \rho =\rho(\hat{T}_b,s_A)-\rho(T_m,0)$ is the density difference between the cooled saltwater flowing downwards below the block at temperature $\hat{T}_b$ and the freshwater in the meltwater layer which is at $T_m$. $\hat{T}_b$ is slightly smaller than $T_b$ and is given by Eq.~(\ref{Eq:Tbhat}).

Because of mass conservation in the meltwater layer, the mass of ice that melts between 0 and $x'$ must equal the mass of meltwater passing through the slice perpendicular to the block at $x'$ at steady state, \textit{i.e.}, 
$$    \rho_i v_m x' = \int_0^{h(x')} \rho_0 w(x',y')\mathrm{d}y'=\frac{g \sin \theta \Delta \rho}{3 \nu} h(x')^3,$$
with $\rho_0=\rho(T_m,0)$. Rearranging, we find
\begin{equation}
    h(x')=\left(\frac{3\nu\rho_i v_m x'}{g \sin\theta\Delta\rho}\right)^{1/3}.
    \label{eq:meltlayerthickness}
\end{equation}
We can discuss the approximations made thus far using this estimate for the meltwater thickness and flow speeds before actually calculating the contribution of the meltwater layer to $U_b$. The velocity of the meltwater current averaged over the melt-layer thickness $\langle w \rangle_{y'} = \frac{2}{3} \, K \, (h(x'))^2$. 
{We have $v_m \,\blue{\approx}\, 40$ $\mu$m $s^{-1}$ at $20\,^\circ$C from Fig.~\ref{MeltingrateSalt}(a),   $\Delta \rho/\rho \approx 0.03$ between freshwater and saltwater with $s_A = 35$ g kg$^{-1}$ at $0\,^\circ$C. Thus, for a block of length $L=200$ mm,  $h(L) \approx 0.5$ mm, $w=\langle w \rangle_{y'} (x'=L) \approx 10$\,mm\,$s^{-1}$ and a Reynolds number $Re=\dfrac{w \, h}{\nu} \approx 5$. 
The assumption that the flow is laminar therefore appears reasonable. 
However, the laminar meltwater flow hypothesis would not be valid for ice blocks that are a few meters in length, because $Re$ would be of order $100$. 
Further, the advection time scale $L/\langle w \rangle_{y'}$ is of order $20$\,s, while the salt diffusion time scale across the melt-layer can be estimated as $h^2/D \approx 250$~s, and the temperature diffusion time scale is $h^2/\kappa \approx 2.5$~s. These estimates are consistent with our assumption that salt does not have time to diffuse into the meltwater layer, which consequently remains as pure water along the length of the ice block. It can be noted that $h$ is smaller than $\delta_T$ by only a factor of 4 at $20^{\circ}$C, which means that neglecting $h$ when calculating $\delta_T$ and assuming a uniform melt layer temperature at $T_m$, are rather crude approximations.}

\subsection{Competition between the meltwater layer flow and the cooled layer flow}
\label{Competition}

\begin{figure}[h]
    \centering
    \includegraphics[width=0.9\textwidth]{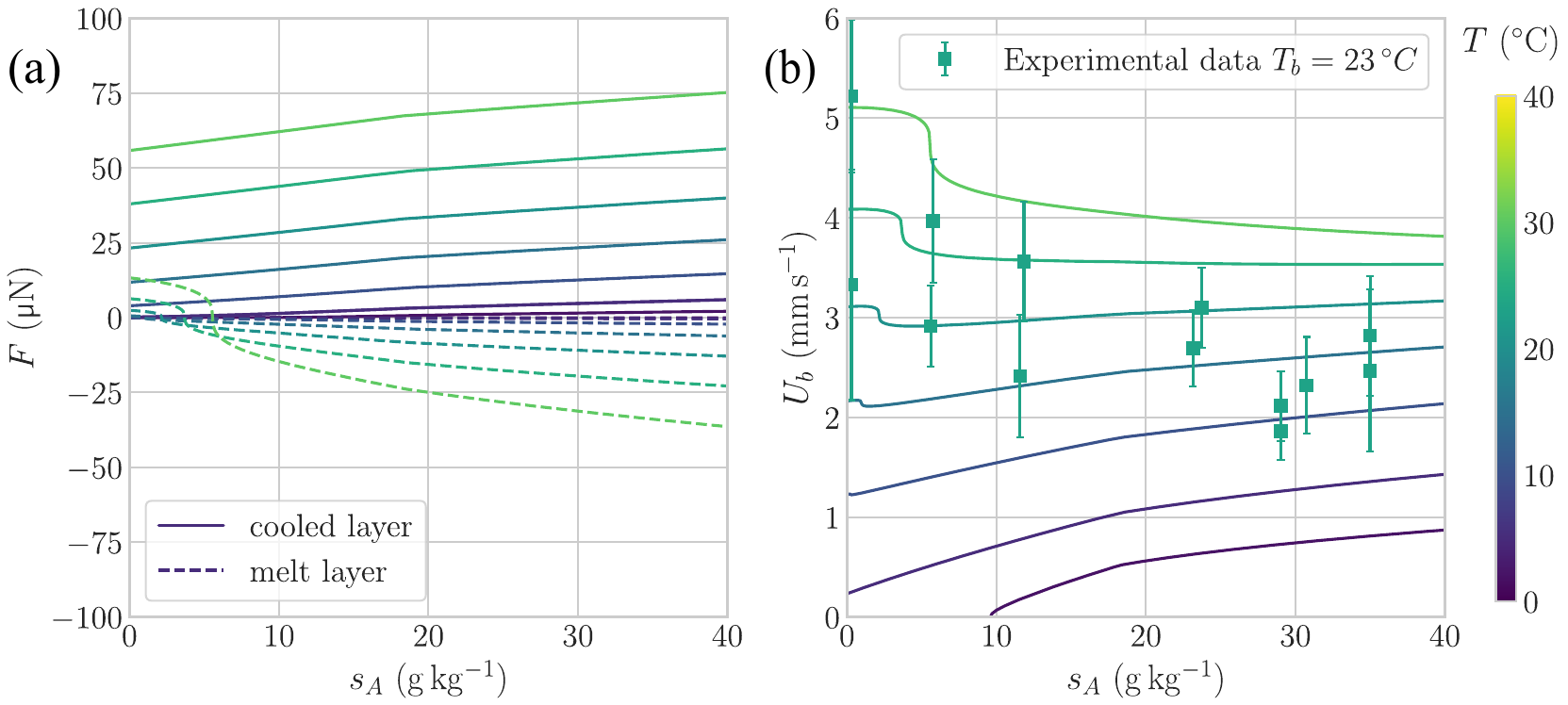}
    \caption{\blue{Theoretical estimation of melting-driven propulsion forces and corresponding terminal velocities in salt water.} (a) The propulsion force in salt water for various bath temperature $T_b$. Continuous line, $F_\mathrm{c}$ contribution of the cooled layer in salt water according to Eq.~(\ref{Fc}). Dashed line, $F_m$ contribution of the melt layer, Eq.~(\ref{Fmelt}). (b) The ice propulsion velocity $U_b$ as a function of bath salinity for an ice block of length 21 cm and inclination $25^\circ$ for different bath temperatures $T_b$, taking into account the contribution of the melt layer. The theoretical estimate is obtained using Eq.~(\ref{UbthS}). \blue{The rescaled measured terminal velocity $\chi_s\, U_b$
     for $T_b=23\,^\circ$ C from Fig.~\ref{fig:salinity} are superimposed. These measurement points are approximately distributed around the theoretical curves for $15$, $20$, and $25\,^\circ$C. The experimental dispersion and uncertainty of measurements are too large for a quantitative comparison with the model.} } 
    \label{UbsaltTheo}
\end{figure}

{Notwithstanding these strong approximations, we estimate the horizontal contribution of the meltwater layer force $F_{\mathrm{m}}$  to the total force exerted by the fluid flow on the ice block. Following similar reasoning as used in obtaining the contribution of the cooled layer using momentum balance, we obtain the force due to the momentum leaving the control volume below the block at $x'=L$ by integrating Eq.~(\ref{eq:meltlayervelocity}), 
\begin{eqnarray}
\mathbf{F}_{\mathrm{m}}&=&-W\int_0^{h(L)}\rho_0 w(L,y')^2 \mathrm{d}y' \, \mathbf{e_{x'}} \,,\\
&=& -\dfrac{8}{15}W\rho_0\left(\dfrac{\Delta\rho g \sin \theta}{2 \rho \nu}\right)^2 h(L)^5 \mathbf{e_{x'}}\,,
\end{eqnarray}
where $\mathbf{e_{x'}}$ is a unit vector along $x'$-axis. After replacing $h(L)$ using Eq.~(\ref{eq:meltlayerthickness}),
the magnitude of the horizontal component of the force in the $x$ direction is thus, 
\begin{equation}
    F_{\mathrm{m}}=-  \dfrac{2\cdot3^{2/3}}{5} W\rho_0\cos\theta\left(\dfrac{\Delta\rho g \sin \theta}{\rho \nu}\right)^{1/3} \left( \frac{\rho_i}{\rho_0}v_m L \right)^{5/3}.
    \label{Fmelt}
\end{equation}}
Fig.~\ref{UbsaltTheo}(a) shows plots of $F_{\mathrm{m}}$ and $F_{\mathrm{c}}$ for various bath temperatures. While $F_{\mathrm{c}}$ systematically increase with salinity over all $T_b$, $F_{\mathrm{m}}$ decreases and notably changes sign. It can be noted that $F_{\mathrm{c}}$ is greater than $F_{\mathrm{m}}$, leading to net propulsion consistent with our experimental observations. Only at very low $s_A$ over a range unexplored in our experiments does the Only at very low $s_A$ over a range unexplored in our experiments does the temperature effect on density outweighs that of salinity.

By balancing the drag with $F_{\mathrm{m}}$ given by Eq.~(\ref{Fmelt}) and $F_{\mathrm{c}}$ given by Eq.~(\ref{Fc}), and rewriting, we obtain the terminal ice block velocity in saltwater, 
\begin{equation}
    U_{b,th} =  \sqrt{\frac{2(F_{\mathrm{c}}+F_{\mathrm{m}})}{C_D \,\rho \, W \, L_A}},\,\,\,\,\, {\rm for} \,\,\,F_c > F_m.
    \label{UbthS}
\end{equation}

Fig.~\ref{UbsaltTheo}(b) shows a plot of $U_b$ calculated with Eq.~(\ref{UbthS}) as a function of salinity at various $T_b$, where it can be noted that $U_b$ is positive over a wide range of $T_b$ for sufficiently large $s_A$. While $U_b$ increases for sufficiently low temperatures, it decreases for $T_b$ above $20\,^\circ$C. \blue{Given the large experimental dispersion, our experiments performed in warm baths with $T_b \approx 23\,^\circ$C are consistent with the predicted trend. The experimental data points are spread around the theoretical curves for $T_b=15$, $T_b=20$ and $T_b=25\,^\circ$ C. We note also that our model relies on several assumptions made}, including the rather strong assumption that the layer of meltwater neither mixes with the layer of cooled water below, nor exerts drag. 

It is important to emphasize that our model is valid only when melting is driven by convective instability. It does not apply when the thermal boundary layer is stably stratified, which occurs when the bath temperature is below the critical temperature for a given salinity, \textit{i.e.} $T_b<T_c(s_A)$. 
Our model cannot thus predict the melting velocity and the block velocity for $T_b=1\,^\circ$C and $s_A\leq13$\,g kg$^{-1}$ as shown in  Fig.~\ref{UbsaltTheo}. We would expect that the layer of cooled water, flowing upward in the $x'$-direction while remaining attached to the block, would generate a reactive force that propels the block in the opposite direction. Then, we would \blue{anticipate} that the layer of cooled water, flowing upwards in the $x'$ direction while remaining attached to the block, would propel the block in the other direction.

\subsection{First-order estimate of the terminal velocity of the ice block in salt water}
\label{scalingsSalt}

To estimate the magnitude of propulsion velocity at ocean salinity, where the contribution from bath cooling dominates, we can apply the previous line of reasoning used in Sec.~\ref{scalings}. 
Because the water density maxima is absent for salinity above $24$\,g~kg$^{-1}$, we also do not consider the density contrast \blue{between the temperature of maximum density} $T_c$ and \blue{bath temperature} $T_b$, but rather between \blue{melting temperature} $T_m$ and \blue{bath temperature} $T_b$. Thus, Eq.~(\ref{Eqscalinglaw}) gives us, 
\begin{equation}
 \label{Eqscalinglawsalt}
U_{b,sc} \sim
  \left(\dfrac{\Delta \rho}{\rho}\right)^{1/3} 
 \left(\dfrac{\rho(T_m)^2}{\rho_b^2} \right)^{1/3}
 \left( \dfrac{{L}}{\delta_T} \right)^{1/3} (\kappa \, g)^{1/3},
\end{equation}
where $\delta_T$ is evaluated using Eq.~(\ref{deltaT2}), and $\dfrac{\Delta \rho}{\rho}=2\,\dfrac{\rho(T_m)-\rho(T_b)}{\rho(T_m)+\rho(T_b)}$ is evaluated using TEOS-10 for $s_A =35$ g kg$^{-1}$~\cite{mcdougall2011getting}. We note that here, as in freshwater, the magnitude of the latent heat is not important to determining the velocity, but only the time duration over which the ice block melts.  

Because we assume that the ice  remains in contact with the fresh meltwater layer, $T_m=0\,^\circ$C, and not the lower value expected if the ice is in direct contact with saltwater. Then, for $T_b \approx 20\,^\circ$C, we find $\Delta \rho/\rho \approx 3\times 10^{-3}$, $\delta_T \approx 1.3$~mm, and $U_{b,sc} \approx 8$~mm\,s$^{-1}$, which overestimates $U_b$ by at least a factor 2 compared to experimental measurements. Nonetheless, Eq.~(\ref{Eqscalinglawsalt}) provides a correct order of magnitude for the block speed in saltwater. If we assume $T_m$ decreases with salinity, $U_{b,sc}\approx 8$\,mm~s$^{-1}$ with $T_m=-1.90\,^\circ$C for a salinity of $35$\,g~kg$^{-1}$, which is the same order of magnitude. Therefore, the decrease in the melting temperature with salinity has a little effect on the ice block velocity in warm water, if the ice melting is driven by thermal convection.

\section{Relevance of self-propulsion to natural icebergs}
\label{Icebergs}
Icebergs found in the oceans are typically on the order of 100 meters in size, irregular shaped and composed of frozen freshwater, formed by the calving of ice sheets in Antarctica and Greenland~\cite{cenedese2023}. The iceberg shapes below the waterline are not known well~\cite{mckenna2005,cenedese2023}, and aside from tabular icebergs, the emerged parts are often asymmetric due to processes involved their formation and deterioration~\cite{romanov2012}. Consequently, there is no strong reason to assume that the immersed shape of icebergs is symmetrical. \blue{Recent measurements using automated underwater vehicles have characterized the immersed topography of ice shelves \cite{schmidt2023heterogeneous,lawrence2023crevasse}. This method could be used to enhance our knowledge of the shapes of free-floating icebergs underwater.}
Icebergs are predominantly reported in seawater with temperatures between $0$ and $5\,^\circ$C~\cite{budd1980antarctic}. As such, our measurements are not conducted under typical conditions in which icebergs are found. However, icebergs have been reported drifting in warmer waters  near Newfoundland (around $45^\circ$ latitude) in spring, where the ocean temperature can reach 10$\,^\circ$C, according to the Arctic Environmental Response Management Application (ERMA) of the National Oceanic and Atmospheric Administration (NOAA), (see \url{https://erma.noaa.gov/arctic}). Since icebergs can drift thousand of kilometers and reach these warm waters, the propulsion mechanism discussed in our study may be relevant to some iceberg scenarios. Using the scaling law for salt water given by Eq.~(\ref{Eqscalinglawsalt}), we estimate the propulsion speed for an iceberg of size about $200$~m melting in a seawater at $10\,^\circ$C, and find $U_b \approx 0.05$ m s$^{-1}$. While smaller, it is of the same order of magnitude as the typical drift velocities of icebergs, which is about $0.1$\,m~s$^{-1}$~\cite{schodlok2006weddell,torbati2020evaluation}. This estimate suggests that while melting-induced propulsion alone cannot fully account for iceberg drift, it may represent a non-negligible contributing factor.
 
Moreover, even though the melting dynamics is not completely characterized for natural icebergs, the presence of currents generated by melting and asymmetric geometry is sufficient to generate a propulsion force, as was also noted \blue{in Chaigne et al.}~\cite{Chaigne2023}. The concentrated rising currents of fresh meltwater near icebergs are well documented in field measurements~\cite{HELLY20111346,YANKOVSKY20141,cenedese2023,nash2024turbulent}. In a recent study~\cite{nash2024turbulent}, coupled water velocity and temperature measurements were performed near a vertical iceberg face approximately 10\,m deep, melting in seawater at 3.6$\,^\circ$C. At a depth of 6.5~m, a strong, rising meltwater plume was observed, with a velocity of approximately 5~cm\,s$^{-1}$ extending 20~cm from the ice wall, in the absence of external currents. The velocity and width of a buoyant plume evolve in $d^{1/3}$ and $d$ respectively~\cite{mcconnochie_kerr_2016}, where $d$ is the vertical distance to plume's starting point, which is the bottom of the iceberg in this case. Using Eq.~(\ref{Fc}), we find therefore at first order $F_c /W \sim \rho_b \times \delta_x \times v_p^2 \,\blue{\approx}\, 1.9$ N m$^{-1}$. For comparisons, a typical wind of velocity $u_W \approx 5$ m s$^{-1}$ acting on a surface $W\times H_e$, with $H_e \approx 1$\,m representing the emerged height, 
the force due to the wind is about $F_w /W \sim \rho_{air} \, \times H_e \times u_W^2 \,\blue{\approx}\, 25$ N m$^{-1}$. In this estimate, the melting induced propulsion is about 7.5\% that the one due to the wind. For a larger iceberg of draft and width $200$ m, a recent model~\cite{mcconnochie_kerr_2016} estimates the plume thickness and velocity to be about $10$\,m and $0.06$\,m s$^{-1}$ for an ocean temperature of $1\,^\circ$C. Similar orders of magnitudes are also obtained for plumes under ice shelves~\cite{Hewitt2020}, which are the glaciers attached to Antarctica and Greenland and give birth to most of the icebergs. Then, using again Eq.~(\ref{Fc}), a first order estimation gives $F_c\,\blue{\approx}\, 7 \times 10^3$\,N, whereas the contribution of the wind of $5$\,m~s$^{-1}$ acting on a surface of order $10 \times 200$ gives $F_w \,\blue{\approx}\, 5\times 10^4$\,N. In this case, the ratio of melting to the wind contribution to propulsion is about 14\%. 

These estimates suggest that while meltwater-driven propulsion is generally weaker than wind forcing, it may still contribute a small but non-negligible amount to iceberg motion. Importantly, because this effect is driven by solutal (compositional) convection, it remains effective even in cold water. The magnitude and direction of this force depend on the iceberg’s asymmetry and the inclination of its submerged surfaces.
We also note that the convection flow generated by heat transfer into the surrounding salt water during melting has not been directly identified in field studies. It is difficult to separate this thermally driven flow from ambient ocean currents, as it lacks associated salinity gradients and may extend over length scales comparable to the iceberg itself. According to our estimates shown in Fig.~\ref{UbsaltTheo}, this thermal contribution could dominate at sea surface temperatures above approximately $5\,^\circ$C.

\section{Discussion and Summary} 
\label{Discussion}
We have experimentally demonstrated that asymmetric floating ice blocks melting in water can self-propel at sufficiently high bath temperatures, extending previous work on propulsion driven by solutal convection~\cite{Chaigne2023}. Our study focused on a simplified asymmetric geometry, in which the convective flow is primarily generated by a planar face inclined at an angle to the horizontal. This design enables quantitative comparison with phenomenological models. Importantly, the propulsion phenomenon observed is robust and not strongly sensitive to the precise dimensions of the ice block, the degree of rounding, or the presence of encapsulated air bubbles. Since the inclination angle remained approximately constant during the first half of the block’s lifetime, we neglected shape changes due to melting. For more complex asymmetric shapes, where the immersed face may exhibit slope variations, the key aspects of the propulsion mechanism should still apply. However, accurately predicting the propulsion velocity in such cases would require numerical simulations of the flow. Further studies exploring propulsion in asymmetric ice blocks with alternative geometries would be a valuable extension of this work.

Moreover, we have shown that the same propulsion mechanism applies for asymmetric ice blocks melting in water bath as the salinity is increased from fresh water to that of the ocean. According to our model presented in Section~\ref{Competition}, the cooling of the bath due to melting and the resulting sinking flow generates a higher force than the one associated with the rising meltwater flow. Consequently, we obtain a propulsion force of the same direction and nearly the same magnitude as in freshwater. Nonetheless, a significant decrease of the terminal velocity is observed when the salinity is increased from zero to that of oceans salinity. This observation was explained at least partially by the force generated by the opposite flow dynamics of the melt layer.

In freshwater baths the propulsion is driven by the detached thermal convection flow, the mechanism is  therefore valid for $T_b$ above the temperature $T_c=3.98\,^\circ$C where the water density maximum. {As shown in Fig.~\ref{UbsaltTheo}, the model remains valid when ice melting occurs in saltwater and is driven by thermal convection, \textit{i.e.} $T_b>T_c(s_A)$. As $T_c(s_A)$ decreases with $s_A$ the validity domain of the propulsion mechanism is extended to lower temperatures as long as $T_b>T_m(s_A=0)=0\,^\circ C$.} However, further experimental tests are required of melting driven propulsion as bath temperatures approach the melting temperature. 

We have also assumed that the meltwater layer remains laminar and does not mix with the surrounding bathwater due to flow turbulence. Under this assumption, the meltwater layer consists entirely of freshwater, and therefore does not alter the melting temperature $T_m$ of the ice block. While this approximation is likely reasonable for centimeter-scale ice blocks, it may break down at larger scales.
Indeed, supported by in situ measurements~\cite{jenkins2010observation,davis2019turbulence}, models of flow around kilometer-scale ice shelves account for a turbulent boundary layer~\cite{Malyarenko2020}. These considerations highlight the need for further experimental investigations of ice melting in water across a broad range of temperatures approaching $T_m$.

In summary, our work has experimentally demonstrated that asymmetric ice blocks melting in both freshwater and saltwater can self-propel due to convection flows generated during melting. In the context of iceberg drift, the primary forces currently considered are ocean currents, wind stress, surface waves, the Coriolis effect, and pressure gradients~\cite{Wagner2017,Marchenko2019,wagner2022winds}, with wind being the dominant factor for icebergs smaller than a few hundred meters~\cite{Wagner2017}. Because the submerged geometries of icebergs are generally uncharacterized, fluid drag coefficients are typically estimated only approximately.
Under typical environmental conditions, our order-of-magnitude estimates suggest that melting-induced propulsion can contribute a significant force, although typically about 10\% of the magnitude of dominant driving forces, such as wind. Further studies should investigate this melting-induced propulsion in natural icebergs, as it represents a potentially relevant mechanism that is currently overlooked in iceberg drift models.

\begin{acknowledgments} We acknowledge Sylvain Courrech du Pont and Philippe Brunet at MSC, Universit\'e Paris Cit\'e for scientific discussions and technical help. M.B. and M. C. are supported by French Research Council project through grant \textit{PhysErosion ANR-22-CE30-0017}.
\end{acknowledgments}

\appendix

\section{Data sets}
\label{Datasets}
\newpage

\thispagestyle{empty}

\includepdf[pages=-,pagecommand={} ,fitpaper=true,offset=0 -20,clip,delta=0 0, scale=1, frame=false]{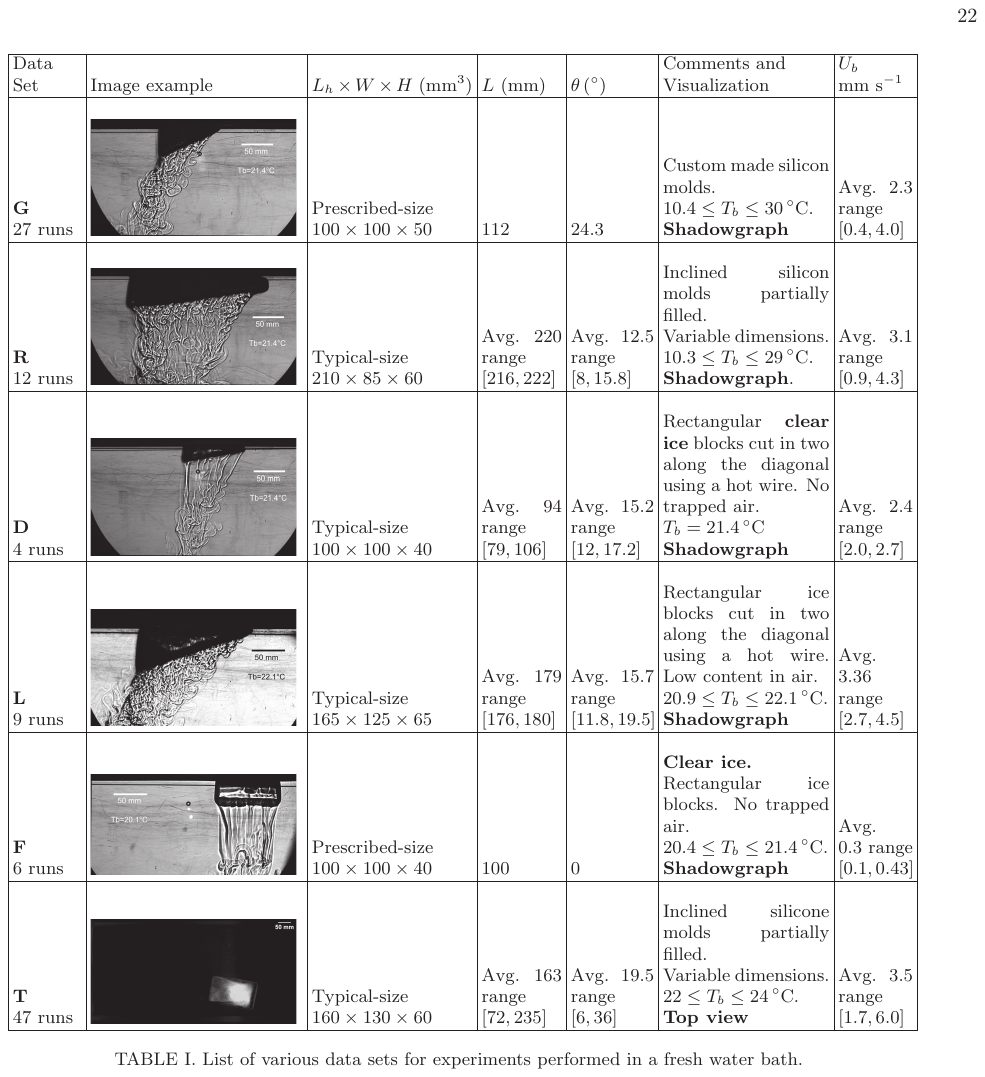}

\newpage

\thispagestyle{empty}

\includepdf[pages=-,pagecommand={} ,fitpaper=true,offset=0 -20,clip,delta=0 0, scale=1, frame=false]{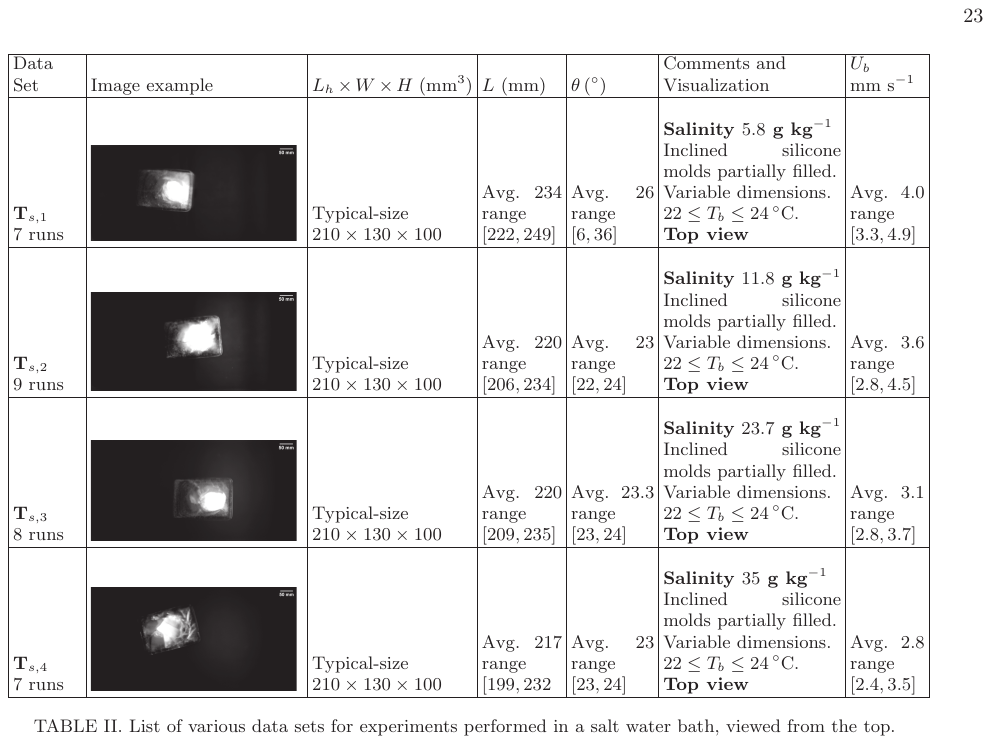}

\newpage

\thispagestyle{empty}

\includepdf[pages=-,pagecommand={} ,fitpaper=true,offset=0 -20,clip,delta=0 0, scale=1, frame=false]{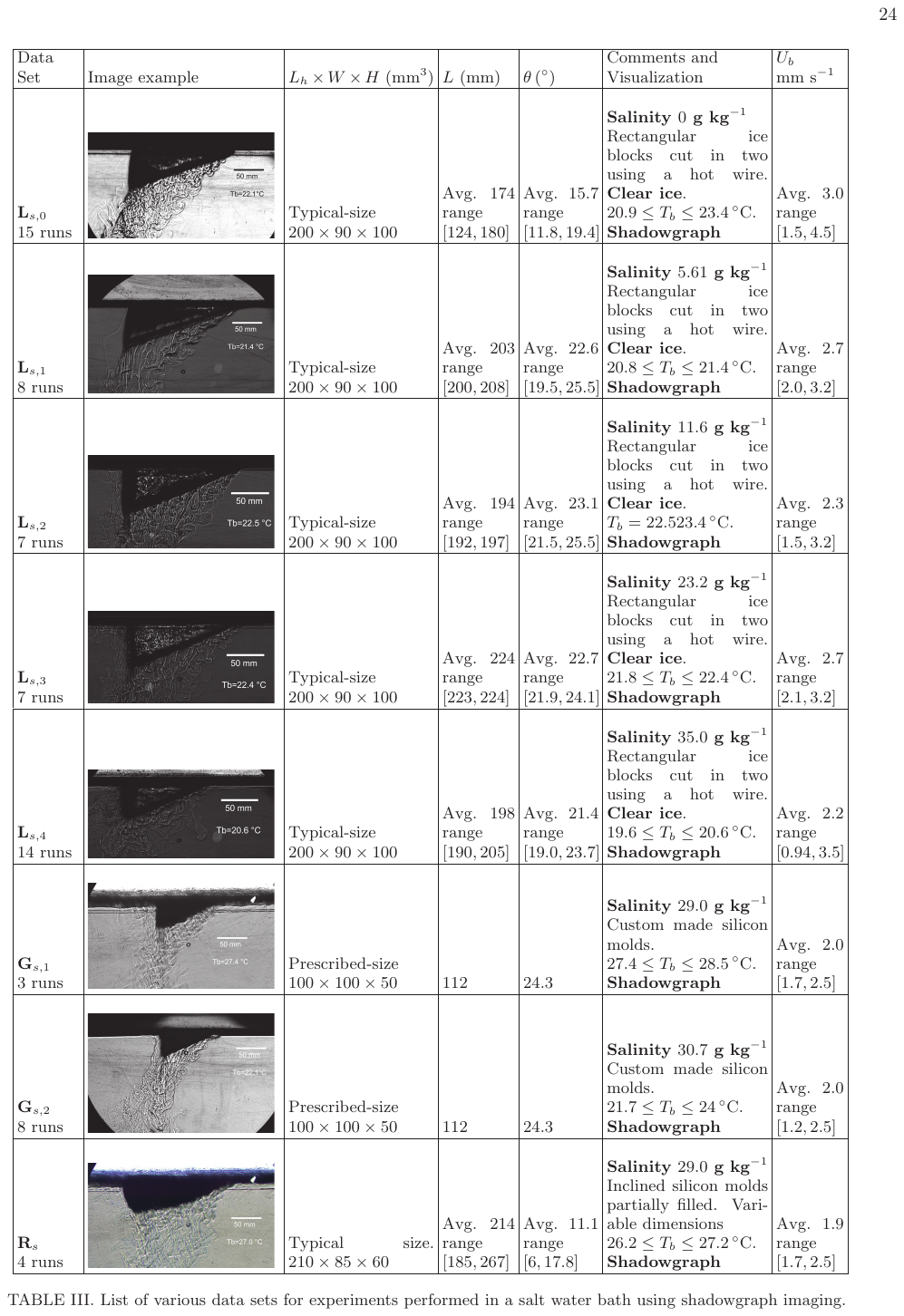}

\clearpage

\section{Stability analysis of wedge blocks}
\label{Sec:StabilityAnalysis}
\begin{figure}[h!]
    \centering
    \includegraphics[width=12cm]{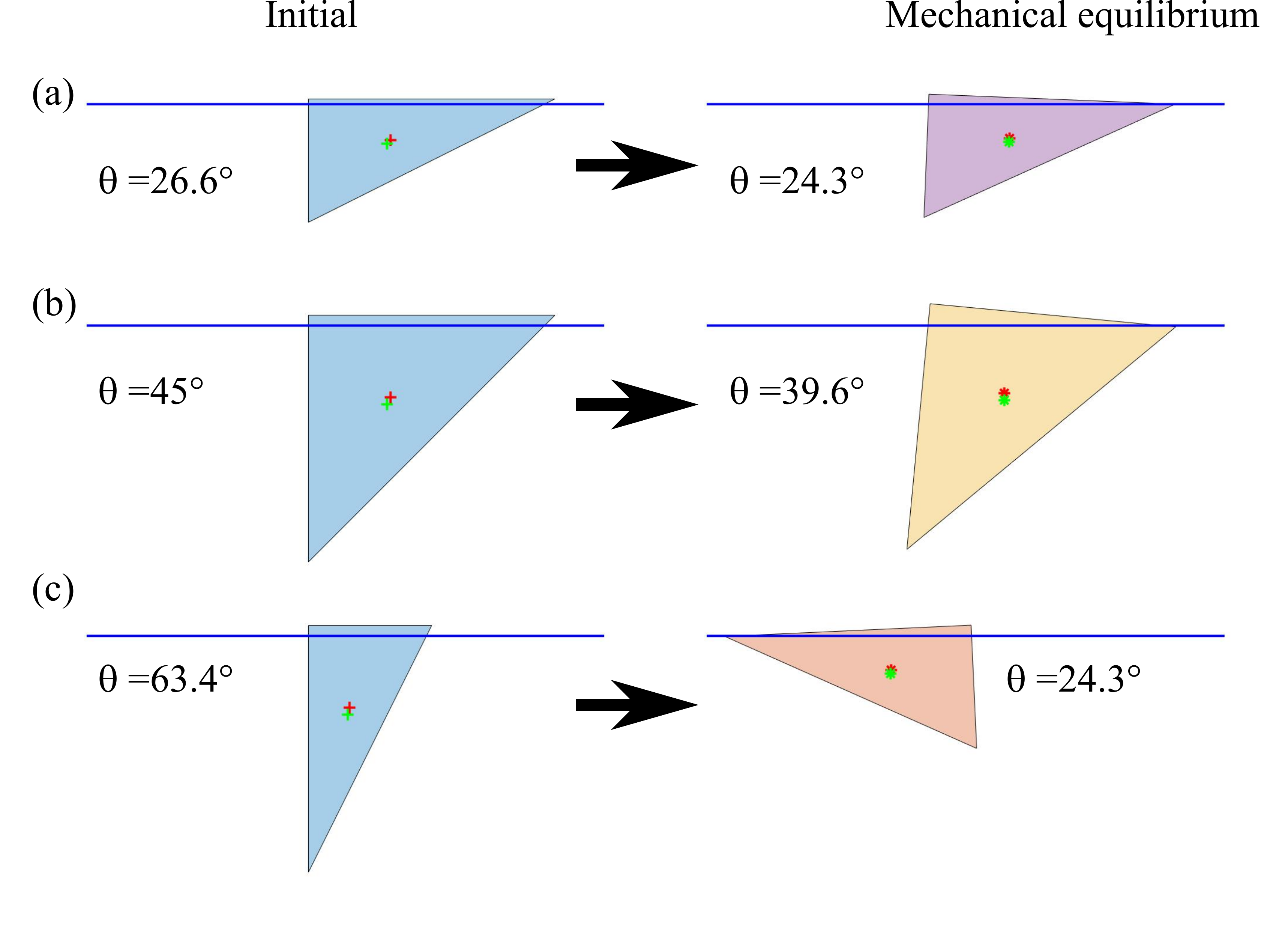}
      \caption{Numerical analysis of stability of wedge ice blocks. (a) $L_h\times H=100 \times 50$ (dimensionless units). 
      (b) $L_h\times H=100 \times 100$. 
      (c) $L_h\times H=50 \times 100$. 
      The initial and final position of the center of mass and buoyancy are denoted with red and green markers, respectively. }    \label{StabilityAnalysis}
\end{figure}

The mechanical equilibrium of our ice blocks is reached after few oscillations over a few seconds when the centers of gravity and buoyancy are vertically aligned. In our right angle triangular prism geometry, the hypotenuse is immersed in the fluid at equilibrium, whereas the second longest side of the right triangle emerges slightly above water as shown for example in Fig.~\ref{fig:ice} (a). To predict orientation of an ice block floating in water, {we use a 2D numerical iterative method. The ice block is assimilated to a planar polygon in the vertical plane. We compute first the center of gravity of the polygon. Then, we find the position of the horizontal free-surface by adjusting its level from the bottom until the surface inside the polygon and below this line is equal to the full polygon surface multiplied by the ratio $\rho_{ice}/\rho_b$. Then, the position of center of buoyancy is computed as the center of mass of the immersed part of the polygon. We start from the initial block orientation where the surface of the longest side is aligned with the water surface. If the center of mass and of buoyancy are not vertically aligned}, a small rotation is applied and the level of the water surface is reevaluated. After few iterations, the final block disposition is reached. Numerically, we find the largest inclination for an isosceles right triangle is of $39.6^\circ$. Therefore, the possible range of the inclination $\theta$ lies between $0^\circ$ and $39.6^\circ$. Few examples of the procedure is illustrated in Fig.~\ref{StabilityAnalysis}. For the example of an ice block with dimensions $100\times 100 \times 50$ mm$^3$, the inclination initially of ${26.6}^\circ$ becomes at mechanical equilibrium $24.3^\circ$. {The wedge ice blocks are considered here at rest in this estimation of their orientation. However, for the drag force torque to counterbalance the buoyancy force torque, \textit{i.e.} $\rho_b \, U_b^2 \gg \rho_{ice}\,g \, L$, translation speeds of the order of meter per second are required.}

\section{{Contribution of the skin friction to the total drag.}}
\label{Skinfriction}

\begin{figure}[h]
    \centering
\includegraphics[width=12cm]{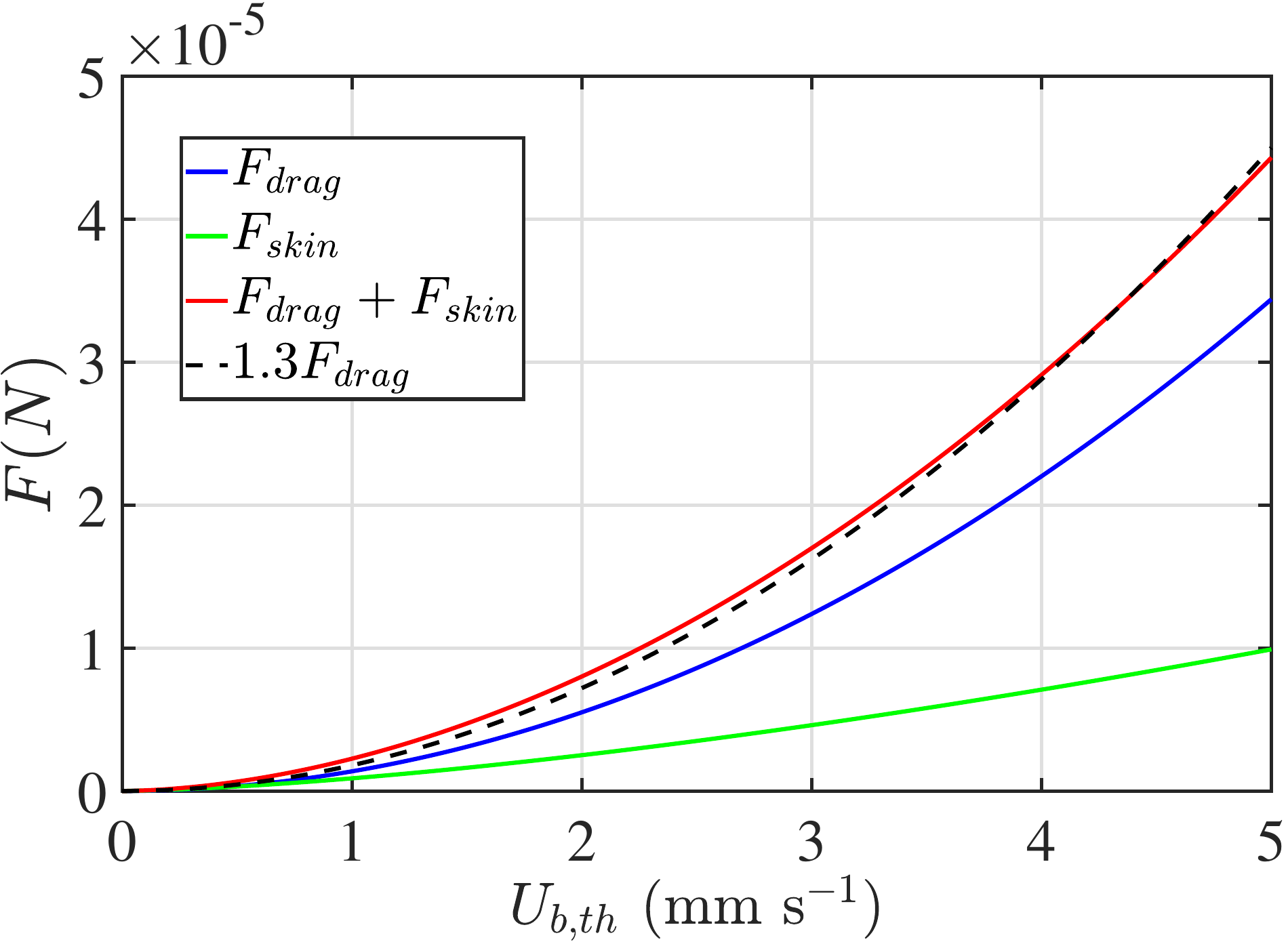}
    \caption{Estimation as a function of the terminal ice block velocity $U_b$ of the contributions of the inertial drag friction $F_{drag}$ and of the skin friction on the lateral sides $F_{skin}$ for a wedge ice block with dimensions $100 \times 50 \times 50$ mm$^3$ (Dataset G). The total drag $F_{drag}+ F_{skin}$ is well approximated by $1.3\, F_{drag}$. The total drag can be considered as an inertial drag friction with an effective coefficient $C_d=0.78\approx 0.8$, instead of $C_d=0.6$ without skin friction.} 
    \label{FdragplusFskin}
\end{figure}

{We estimate the contribution of the skin friction to the total drag experienced by the wedge ice block moving at terminal velocity. This effect was neglected for inclined candy boats~\cite{Chaigne2023}, because the surface of their lateral sides is small compared to their frontal surface. Yet, this hypothesis is less valid for the wedge ice blocks. We estimate the magnitude of skin drag for an ice wedge block moving at its terminal velocity and of dimensions $100 \times 50 \times 50$ mm (Dataset G).
The inertial drag reads,
\begin{equation}
    F_{drag}=\frac{1}{2}\,C_d\,\rho_b\,L\, \sin \theta\,W\, U_b^2 \,, \end{equation}
whereas the skin friction is evaluated by computing the friction force for a laminar velocity boundary layer on a plate (Blasius profile)~\citep{GuyonHulinPetit}, \textit{i.e.}, 
\begin{equation} F_{skin}=2\,\frac{4}{3}\,\rho_b\,U_b^{3/2}\,\nu^{1/2}\,\int_0^{H}\,l_h(y)\,d_y=\frac{8}{9}\,\rho_b\,U_b^{3/2}\,\nu^{1/2}\,H\,L_h \,.
\end{equation}
The factor $2$ accounts for the two lateral sides and $l_h(y)$ is the length of the horizontal section of the block along $x$ at a given vertical coordinate $y$. Taking the value $C_d=0.6$, we find $F_{drag} > 2 F_{skin}$ for $U_b > 1.5$ mm s$^{-1}$. We compute and plot the total drag $F_{drag}+F_{skin}$ as a function of the terminal velocity $U_b$ in Fig.~\ref{FdragplusFskin} and observe that it is well fitted by $U_b^2$ over the range $[1,5]$~mm\,s$^{-1}$, \textit{i.e.} $F_{drag}+F_{skin} \approx 1.3 \, F_{drag}$. Therefore, the effect of the skin drag in our experiment increases the overall inertial drag friction coefficient, resulting in $C_d=0.78$ instead of $C_d=0.6$. Consequently, we choose to consider only an inertial drag in our model, allowing a simple relation between the terminal velocity and the propulsion force, when in fact the drag coefficient is an effective coefficient. 
This conclusion would be truer for larger or faster ice blocks which move at higher Reynolds number. In that case, the skin drag can be expected to be negligible in comparison with the inertial drag.}

\end{document}